\documentclass[12pt,oneside]{article}
 
\renewcommand{\arraystretch}{1.2}
\usepackage[latin1]{inputenc}
\def \mb{\mathversion{bold}}
\def \mbb{\mathbb}
\def \ds{\displaystyle}

\def \scrs{\scriptstyle}
\def \MF{Mend\`es France}
\def \varep{\varepsilon}
\def \dist{\rm dist}
\def \supp{{\rm supp}}
\def \ol{\overline}

\def \wt{\widetilde}

\def \gs{\geqslant}
\def \ls{\leqslant}
\newtheorem{theo}{Theorem}[section]
\newtheorem{pro}{Proposition}[section]
\newtheorem{cor}{Corollary}[section]

\newtheorem{lemma}{Lemma}[section]
\newtheorem{rem}{Remark}[section]

\def\qed{\hfill {\mb $\Box$}}
\def\prf{{\bf Proof.}}

\usepackage{latexsym}
\usepackage{amstext}
\usepackage{amssymb}
\usepackage{graphicx}
\usepackage{color}
\usepackage{amsmath}
\usepackage{amsfonts}

\date{}

\title{\large \bf MEND\`ES FRANCE AND THERMODYNAMICAL SPECTRA: A COMPARATIVE STUDY OF CONTRACTIVE AND EXPANSIVE \\ FRACTAL PROCESSES}

\author{ROBERTA HANSEN$^*$\  and \ MAR\'IA N. PIACQUADIO$^\dag$\\ $^*$\it{Departamento de Matem\'atica,}\\ \it{Facultad de Ingenier\'{\i}a, Universidad de Buenos Aires,}\\ \it{Avenida Paseo Col\'on 850, CP: 1063, Buenos Aires, Argentina.}\\ \it{rhansen@fi.uba.ar}\\$^\dag$\it{Secretar\'{\i}a de Investigaci\'on y Doctorado,} \\ \it{Facultad de Ingenier\'{\i}a, Universidad de Buenos Aires.}\\ \it{gdutt@mgminter.com}}

\begin{document}

\maketitle

{\it Nature exhibits not simply a higher degree but an altogether different level of complexity. The number of distinct scales of length of natural patterns is for all practical purposes infinite.}
\medskip

\hfill{Benoit B. Mandelbrot \cite{Man82}}

\begin{abstract}
This paper presents a comparative study of two families of curves in $\Bbb{R}^n$. The first ones comprise self--similar bounded fractals obtained by contractive processes, and have a non--integer Hausdorff dimension. The second ones are unbounded, locally rectifiable, locally smooth, obtained by expansive processes, and characterized by a fractional dimension defined by M. Mend\`es France. We present a way to relate the two types of curves and their respective non--integer dimensions. Thus, to one fractal bounded curve we associate, at first, a finite range of Mend\`es France dimensions, identifying the minimal and the maximal ones. Later, we show that this discrete spectrum can be made continuous, allowing it to be compared with some other multifractal spectra encountered in the literature. We discuss the corresponding physical interpretations.
\end{abstract}

{\bf keywords:} Mend\`es France and Hausdorff dimensions, self similar curves, expanded curves, multifractal spectra.

\section{Introduction}
Fractal geometry focuses on the local non--smoothness of physical and mathematical objects or sets, by means of their fractal dimension, of which there are several formulations: box dimension, Hausdorff, etc. These dimensions represent the ``amount of space" occupied by the set, measuring its degree of ``wrinkledness", when viewed at smaller and smaller scales. Within this fractal universe we will restrict ourselves to bounded self--similar curves $F$ in $\mbb{R}^n$, of Hausdorff dimension, $\dim_H(F)$, larger than unity. The fact that $\dim_H(F)\!>\!1$ indicates that $F$ has infinite length, and therefore, confined to its convex hull, it has to be infinitely folded or wrinkled, in a self--similar way. This characterizes the first family of curves which we call ${\cal F}_H$.
\smallskip

The second family of curves we deal with in this paper, ${\cal F}_{MF}$, are unbounded, and free to spread their infinite length all around the $n$--dimensional space. Each such curve $\Gamma$ is locally rectifiable and locally smooth, and hence with $\dim_H(\Gamma)\!=\!1$. M. \MF\ defined a fractional dimension $\dim_{MF}$ for them \cite{MF}. This dimension ``looks at" such a curve $\Gamma$ from afar, further and further away, instead of ``at smaller and smaller scales", as mentioned above. The idea is to ``zoom out" and study the scaling properties of their lengths when growing to infinity. 
\smallskip

Curves $F$ in ${\cal F}_H$ are obtained by means of contractive processes and contractive ratios, we will obtain curves $\Gamma$ in ${\cal F}_{MF}$ by expansive ones.
\smallskip

It is worth commenting that this approach is completely different from Strichartz's \cite{Stz} \emph{reverse iterated function system} which constructs a new limit fractal set with the same dimension as the original. Our approach is to ascribe, to each $F$, a gamut of unbounded curves, locally smooth, together with their corresponding \MF\  dimensions. 
\medskip

The paper is organized as follows: In Sec. 2 we recall the definition of self--similar bounded curves $F$ in ${\cal F}_H$ by means of their finite number of contractive ratios $a_i$, $i\!=\!1\hdots,N$; in Sec. 3 we do the same with the \MF\ dimension via the expansive ratios $1/a_i$; in Sec. 4 we construct, for a given $F$, a curve $\Gamma^{a_i}$ in ${\cal F}_{MF}$ for each $a_i$, i.e. we ascribe to $F$ a finite spectrum of $\dim_{MF}(\Gamma^{a_i})$. We identify the maximal and minimal dimensions of this spectrum as the Hausdorff and the divider dimensions of $F$. We test the sensitivity of $\dim_{MF}$: a minute change in the value of $a_i$ implies a variation in the dimension  $\dim_{MF}(\Gamma^{a_i})$. In Sec. 5 we make continuous the discrete spectrum obtained in Sec. 4. In Sec. 6 we compare this continuous multidimensional or multifractal MF spectrum with three multifractal spectra encountered in the literature: a) the spectrum of R\'enyi generalized dimensions, b) the thermodynamical formalism $(\alpha,f(\alpha))$ with an appropriate measure, and c) the corresponding range $[\alpha_{\min},\alpha_{\max}]$. Sec. 7 summarizes the conclusions.

\section{Curves constructed by similarities}
Let $A\!=\!A_1,A_2,\hdots,A_{N+1}\!=\!B$ be $N+1$ different points in $\mbb{R}^n$, satisfying $\dist(A_i,A_{i+1})\!<\!\dist(A,B)\!=\!1$, for all $i\!=\!1,\hdots,N$. Let $S_i\!:\!\mbb{R}^n\!\to\!\mbb{R}^n$ be $N$ similarities such that $S_i(AB)\!=\!A_iA_{i+1}$, so $S_i$ are contractions, and $a_i\!=\!\dist(A_i,A_{i+1})\!<\!1$ their ratios of similarities or \emph{contractors}, $1\!\ls\!i\!\ls\!N$.

\begin{figure}[h]                      
\includegraphics[width=14.5cm,height=10cm]{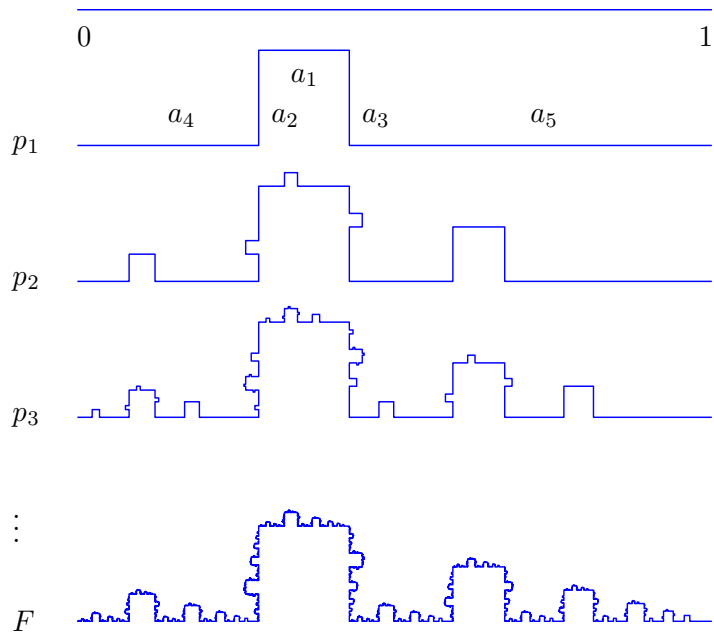}      
\vspace{-1cm}                          
\caption{\it Self--similar curve $F$ in $\mbb{R}^2$; generatrix $p_1$ with $a_1\!=\!a_2\!=\!a_3\!=\!1/7$, $a_4\!=\!2/7$ and $a_5\!=\!4/7$.}                           
\label{fractal-autosem}                
\end{figure}

Let $p_1$ be the polygonal whose vertices are $A_1,A_2,\hdots,A_{N+1}$, so it is formed by the $N$ segments $A_iA_{i+1}$; this is the first polygonal approximation of the curve $F$. We call $p_1$ the \emph{generatrix} of $F$. The polygonal $p_2$ is obtained by replacing each segment $A_iA_{i+1}$ by its copy $S_i(p_1)$, which has the same endpoints $A_i$ and $A_{i+1}$. So $p_2$ has $N^2$ segments, and $p_2\!=\!\bigcup^N_{i=1}S_i(p_1)$, and so forth. Assuming the polygonal $p_{k-1}$ has been constructed, we replace the segment $A_iA_{i+1}$ by $S_i(p_{k-1})$, obtaining a polygonal $p_k$ made of $N^k$ segments, such that $p_k\!=\!\bigcup^N_{i=1}S_i(p_{k-1})$. It is proved in \cite{CT} that the sequence of \emph{prefractals} $\{p_k\}$ converges, as $k\!\to\!\infty$, according to the Hausdorff distance, to a limit curve $F$, satisfying
$$F=\bigcup_{i=1}^N S_i(F)\ ,$$
so $F$ is invariant for the iterated function system (IFS) $S_1,\hdots,S_N$, it is infinitely wrinkled, and has self--similar structure. The well--known von Koch curve (where $a_i\!=\!1/3$, $i\!=\!1,2,3,4$), or the curve in Fig. \ref{fractal-autosem}, are examples in $\mbb{R}^2$.

Recall that an IFS $S_1,\hdots,S_N$ satisfies the \emph{open set condition} (OSC) \cite{Fal90,Fal97} if there exists a non--empty bounded open set $U\!\subset\!\mbb{R}^n$ such that
$$\bigcup_{i=1}^NS_i(U)\subset U\ ,$$
with this union disjoint. This criterion guarantees that the components $S_i(F)$ do not overlap ``too much".
\medskip

In what follows, we will call ${\cal F}_H$ the family of self--similar bounded fractal curves, $F$, constructed as described above, and satisfying the OSC. Under these hypotheses, it is known that the Hausdorff dimension of $F$ is the unique value $d\!=\!\dim_H(F)$ that satisfies the \emph{similarity equation} \cite{Fal90}
$$\sum\limits^N_{i=1}a_i^d=1\ .$$

\section{The \MF\ Dimension of Expanded \\ Curves in \mb{$\mbb{R}^n$}}

Let ${\cal F}_{MF}$ be the family of curves $\Gamma\!\subset\!\mbb{R}^n$ that are unbounded, locally rectifiable, and locally smooth, i.e., any arc of $\Gamma$ has finite length. We will give an idea of the ``fractional dimension" defined by \MF\ for this type of curves \cite{MF}.

For a curve $\Gamma\!\in\!{\cal F}_{MF}$, we fix an origin and  consider the first portion $\Gamma_L$ of $\Gamma$ of length $L$.
Let $\varep\!>\!0$ be given and let $\Gamma_L(\varep)$ be the $\varep$--\emph{parallel body} of $\Gamma_L$, also known as the $\varep$--\emph{Minkowski sausage} of $\Gamma_L$ 
$$\Gamma_L(\varep)=\bigcup_{x\in\Gamma_L}B(x,\varepsilon)=\big\{ x\in\Bbb{R}^n /\ {\dist}(x,\Gamma_L)<\varep \big\}\ .$$
Let $\Delta_L$ be the diameter of the convex hull of $\Gamma_L$. Then, the \MF\ dimension of a curve $\Gamma$ is, by definition
\begin{equation}
\label{dimMF-n}
\dim_{MF}(\Gamma)=\lim_{\varep\searrow 0} \liminf_{L\nearrow \infty}\frac{\log \mu^n(\Gamma_L(\varep))}{\log\Delta_L}\ ,
\end{equation}
where $\mu^n(\Gamma_L(\varep))$ denotes the $n$--dimensional volume of $\Gamma_L(\varep)$. It can be proved that the limit, when it exists, has a value between 1 and $n$, and that it {\bf does not} depend on $\varep$, so we can drop it and rewrite (\ref{dimMF-n}) as
\begin{equation}
\label{dimmf}
\dim_{MF}(\Gamma)=\liminf_{L\nearrow \infty}\frac{\log \mu^n(\Gamma_L(\varep))}{\log \Delta_L }\ .
\end{equation}
This remark is very important, because, intuitively, it says that it does not matter how ``fat"  the $\varep$--Minkowski sausage is, but how the sausage ``fills up" the space according to the development of $\Gamma_L$ when $L$ grows. Therefore, we are dealing with a concept of dimension which does not look at the curve at small scales, as the Hausdorff or box--counting dimensions do; on the contrary, this dimension ``zooms out", looking from afar at the behavior of the curve when its length tends to infinity.
\medskip

Notice that, when the $\limsup_{L\nearrow\infty}$ exists and equals the $\liminf_{L\nearrow\infty}$ in Eq. (\ref{dimmf}), it suffices to consider a growing sequence $L_k$, such that $\Delta_{L_{k+1}}\!\gs\!c\,\Delta_{L_k}$, for any constant $c\!>\!1$; in particular $\Delta_{L_k}\!=\!c^k$. 

\section{The Discrete Spectrum of \MF \\ Dimensions}

The two families, ${\cal F}_H$ and ${\cal F}_{MF}$, have no curve in common since their curves have absolutely different geometric features; however, we will make a geometrical process that allows us to link curves of both families, and thereby to relate their respective dimensions. We briefly review the main concepts, geometrical ideas and theorems, given previously in a detailed form in \cite{H-P1}.
\medskip

To start, let us consider a strict self--similar $F\!\in\!{\cal F}_H$, where the generatrix $p_1$ is made of by $N$ segments of equal length $a$, $0\!<\!a\!<\!1$. 
\begin{figure}[t]                                    
\vspace{-0.4cm}                                      
\hspace{-1cm}                                        
\includegraphics[width=17cm,height=11cm]{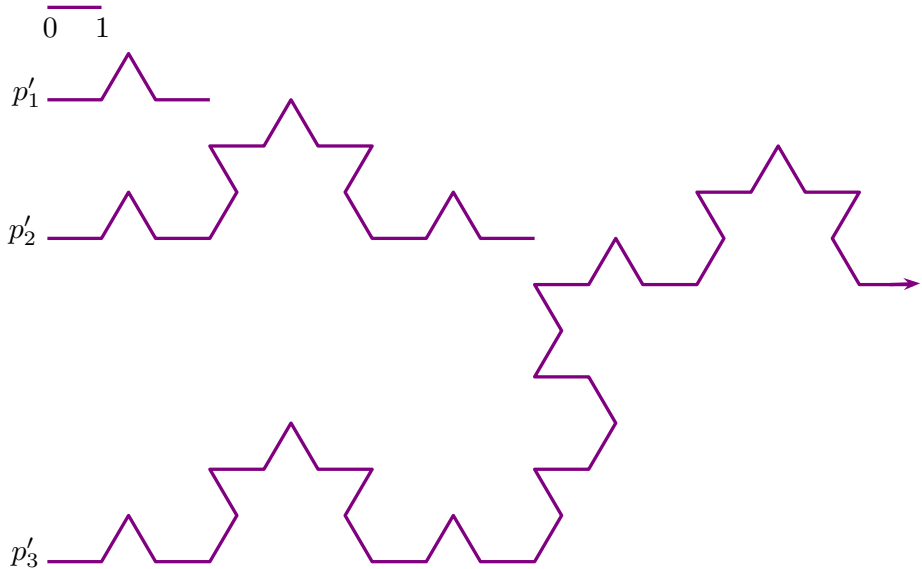}
\vspace*{-1.5cm}                                     
\caption{\it Construction of the expanded von Koch curve.}
\label{Koch-expand}                                  
\end{figure}                                         
For example, in the von Koch curve, $N\!=\!4$ and $a\!=\!1/3$. In the first iteration we construct $p_1'$, identical in shape to $p_1$ but all segments having unit length: $p'_1$ is $p_1$, expanded by a factor of $1/a\!=\!3$. Next, $p'_2$ is $p_2$ expanded by $1/a^2\!=\!3^2$, and so forth, as indicated in Fig. \ref{Koch-expand}. Each $p'_{k+1}$ contains $p'_k$: there is a process of inheritance which guarantees the existence of a limit curve $\Gamma$, continuous, locally rectifiable and locally smooth, unbounded, i.e. in ${\cal F}_{MF}$, that is the ``expanded" and ``unwrinkled" version of $F$. Since all segments in $p'_k$ are of unit length, it is easy to see that $\mu^n(p_k(\varep))\!\approx\!\varep N^k$, and so
$$\dim_{MF}(\Gamma)=\lim_{k\to\infty}\frac{\log \mu^n(p_k(\varep))}{\log \Delta_k}=\lim_{k\to\infty}\frac{\log(\varep N^k )}{\log 1/a^k}=\frac{\ \ \log N}{-\log a}=\dim_H(F)\ .$$
 
\begin{figure}[h]                                 
\hspace{-2cm}                                     
\includegraphics[width=18cm,height=12cm]{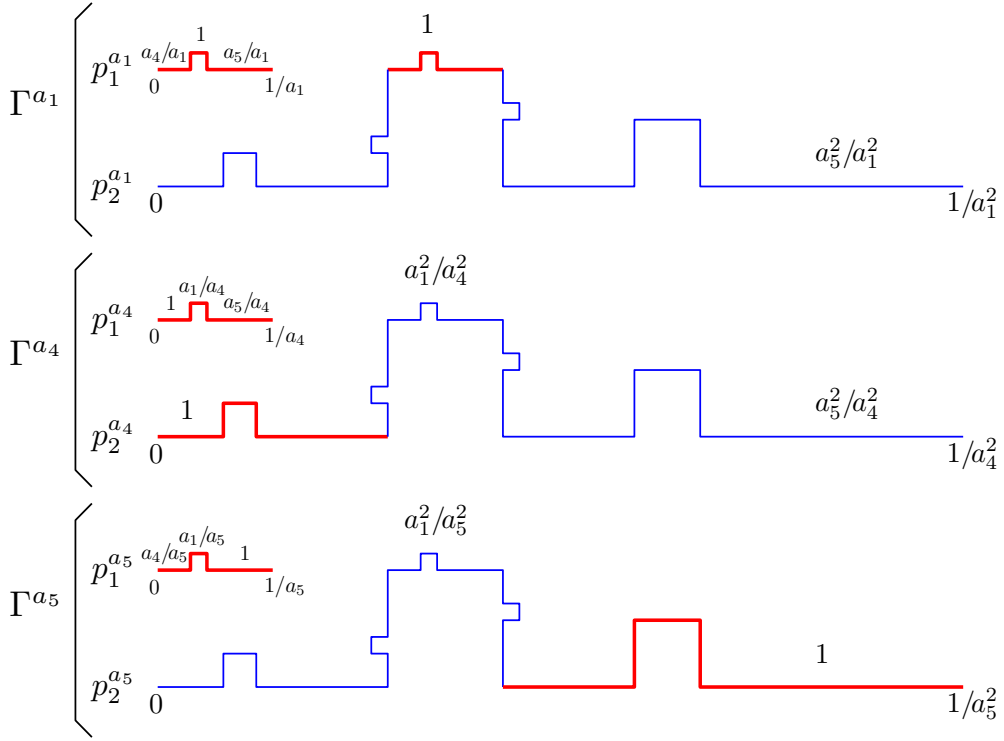}
\vspace*{-1cm}                                    
\caption{\it The first two steps (out of scale) of the construction of the limit curves $\Gamma^{a_1},\Gamma^{a_4}$ and $\Gamma^{a_5}$ as expanded versions of $F$ in Fig. \ref{fractal-autosem}. Note that $p^{a_i}_2$ inherits $p^{a_i}_1$ (red) in each case.}
\label{pol3}                                      
\end{figure}                                      

In the general case of some different values, $a_1\!\ls\!a_2\!\ls\!\cdots\!\ls\!a_N$, as in the example of Fig. \ref{fractal-autosem}, we can enlarge by $1/a_1,\hdots,1/a_N$ expansive factors or \emph{expansors}, and obtain up to $N$ $\Gamma$'s in ${\cal F}_{MF}$, with different values of $\dim_{MF}$. Let us focus on Fig. \ref{pol3}, the largest expansors are $1/a_1, 1/a_2$ and $1/a_3$. Take $1/a_1$ for instance, and observe that in each polygonal $p^{a_1}_k$ the shortest segment has unit length, and that $p^{a_1}_k$ adds new segments, longer than those in $p^{a_1}_{k-1}$: so, in the limit curve $\Gamma^{a_1}$, all segments are longer than or equal to unity. If we take the smallest expansor $1/a_N$ ($1/a_5$ in the same figure) we observe that in each $p^{a_N}_k$ the longest segment has unit length, and that $p^{a_N}_k$ adds new segments, shorter than the ones in $p^{a_1}_{k-1}$. So, the limit curve $\Gamma^{a_N}$ has arbitrarily small segments.

Finally, an intermediate expansor $1/a_i$, $i\!\neq\!1,N$, like $1/a_4$ in the same example, yields polygonals $p^{a_i}_k$ with some segments smaller and others longer than those in $p^{a_i}_{k-1}$, so $\Gamma^{a_i}$ has arbitrarily small and large segments. We have the following result \cite{H-P1}

{\theo\label{dimMF-desig} \quad Let $F\!\in\!{\cal F}_H$ with contractors $0\!<\!a_1\!\ls\!a_2\!\ls\!\cdots\!\ls\!a_N\!<\!1$; the reciprocals $1/a_1\!\gs\!1/a_2\!\gs\!\cdots\!\gs\!1/a_N\!>\!1$ are the expansors constructing limit curves $\Gamma^{a_1},\hdots,\Gamma^{a_N}\!\in\!{\cal F}_{MF}$ respectively. Then
$$ \dim_{MF}(\Gamma^{a_1})\ls \dim_{MF}(\Gamma^{a_2})\ls\cdots\ls \dim_{MF}(\Gamma^{a_N})\ .$$}
We will call $\{\dim_{MF}(\Gamma^{a_1}),\hdots,\dim_{MF}(\Gamma^{a_N})\}$ the \emph{discrete MF spectrum} associated with $F$.

\subsection{Identification of minimal and maximal \mb{$\dim_{MF}$}}
We have also the following results \cite{H-P1}

{\theo \quad
\label{dim-igual} Let $F\!\in\!{\cal F}_H$, and $\Gamma^{a_N}\!\in{\cal F}_{MF}$ the limit curve obtained by the smallest expansor $1/a_N$. Then 
$$\dim_{MF}(\Gamma^{a_N})=\dim_H(F)\ .$$}
{\theo\quad Let $F\!\in\!{\cal F}_H$, and $\Gamma^{a_1}\!\in{\cal F}_{MF}$ the limit curve obtained by the largest expansor $1/a_1$. Then
\begin{equation}
\label{resolv}
\dim_{MF}(\Gamma^{a_1})= 1+ \dfrac{\log\!\left(\!\sum\limits^N_{i=1}a_i\!\right)}{\log\dfrac{1}{a_1}}\ .
\end{equation}}
We have identified the maximal $\dim_{MF}$ of the discrete MF spectrum as the Hausdorff dimension of $F$. To identify the minimal one, let us recall the \emph{divider dimension} or \emph{compass dimension} of $F$: given $\varep\!>\!0$, $M_\varep(F)$ is the maximal number of points $x_0,x_1,\hdots,x_m$ in $F$ (in that order) such that $|x_j\!-\!x_{j-1}|\!=\!\varep$, for $j\!=\!1,\hdots,m$. If $L_\varep$ is the length of polygonal $p_\varep$ that joins the $x_j$, then $L_\varep\!\approx\!\varep\,(M_\varep(F)\!-\!1)\!\approx\!\varep\,M_\varep(F)$. Then, by definition of $\dim_{div}$ we have \cite{Fal90}
\begin{equation}
\label{dimension-div-2}
\dim_{div}(F)=\lim_{\varep\to 0}\frac{\log M_{\varep}(F)}{-\log \varep} =\lim_{\varep\to 0}\ 1+\frac{\log L_\varep}{\log\dfrac{1}{\varep}}\ .
\end{equation}
Next, take the polygonals $p_k$ which yield $F$ and apply the $\varep$--dividing method, but to $p_k$ (instead of $F$) with $\varep_k\!=\!a_1^k$ in each step $k$ ($\varep_k\!\searrow\!0$ when $k\to \infty$, for $a_1\!<\!1$). The length of $p_k$ is $L^k$, where $L\!=\!\sum^N_{i=1}a_i$ is the length of the generatrix $p_1$. Since $a_1$ is the smallest of all ratios, then  $M_{\varep_k}(p_k)\!\approx\!L^k/a_1^k$ which implies, by Eq. (\ref{dimension-div-2}), that
$$\lim_{\varep_k\to 0}\frac{\log M_{\varep_k}(p_k)}{-\log \varep_k}=\lim_{k\to \infty}\frac{\log(L^k/a_1^k)}{-\log a_1^k}=1+ \dfrac{\log L}{\log\dfrac{1}{a_1}}\ ,$$
which, together with Eq. (\ref{resolv}), yield
{\pro\quad Under the same hypotheses of previous theorems, we have $\dim_{MF}(\Gamma^{a_1})$ identified with the concept of $\dim_{div}(F)$.}

\subsection{Sensitivity of \mb{$\dim_{MF}$}}
The following result \cite{H-P1}
{\theo\quad Let $F\!\in\!{\cal F}_{MF}$, and $a_1\!<\!a_i$, $i\!\neq\!1$ then 
$$\dim_{MF}(\Gamma^{a_1})<\dim_{MF}(\Gamma^{a_i})\ ,$$}
implies that, should, e.g. $a_1$ and, say, $a_2$ be infinitely closer, still the \MF\ dimensions of $\Gamma^{a_1}$ and $\Gamma^{a_2}$ would differ. In other words, $\Gamma^{a_1}$ has only segments larger than unity and arbitrarily large, $a_2\!>\!a_1$ implies: some small and smaller segments will be introduced in $\Gamma^{a_2}$. Should these ``wrinkles" be arbitrarily small and difficult to ``see", still they would increase the $\dim_{MF}$.

\section{`Continuization' of the discrete MF spectrum}

\begin{figure}[h]                                       
\vspace{-0.3cm}                                         
\hspace{-3cm}                                           
\includegraphics[width=20cm,height=13cm]{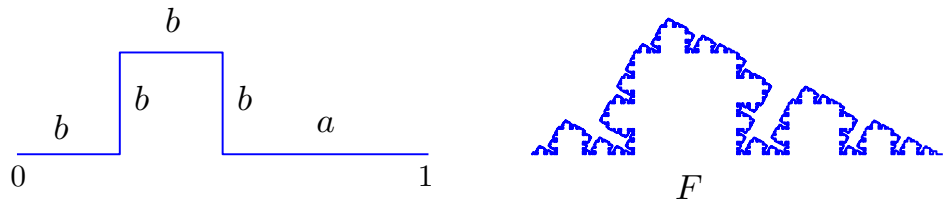}
\vspace{-9cm}                                           
\caption{\it Generatrix $p_1$ with $b\!=\!1/4$, $a\!=\!1/2$ and the yielding fractal curve $F$.}
\label{fractal-ab}                                      
\end{figure}

Indeed, the discrete MF spectrum $\{\dim_{MF}(\Gamma^{a_1}),\hdots,\dim_{MF}(\Gamma^{a_N})\}$ can be made continuous, and in order to demonstrate it, we start with a simple example: only two contractors, $0\!<\!b\!<\!a\!<\!1$ as in Fig. \ref{fractal-ab} (where $b\!=\!1/4$  and $a\!=\!1/2$). In this case, the spectrum has only two dimensions $\{\dim_{MF}(\Gamma^b),\dim_{MF}(\Gamma^a)\}$, with $\dim_{MF}(\Gamma^b)\!<\!\dim_{MF}(\Gamma^a)$. $\Gamma^b$ is the very stretched version, so that it has all segments larger than or equal to unity; whereas $\Gamma^a$, although expanded, is a more wrinkled version since it has segments arbitrarily small.
\begin{figure}[h]                                         
\vspace{-0.3cm}                                           
\hspace{-3cm}                                             
\includegraphics[width=20cm,height=13cm]{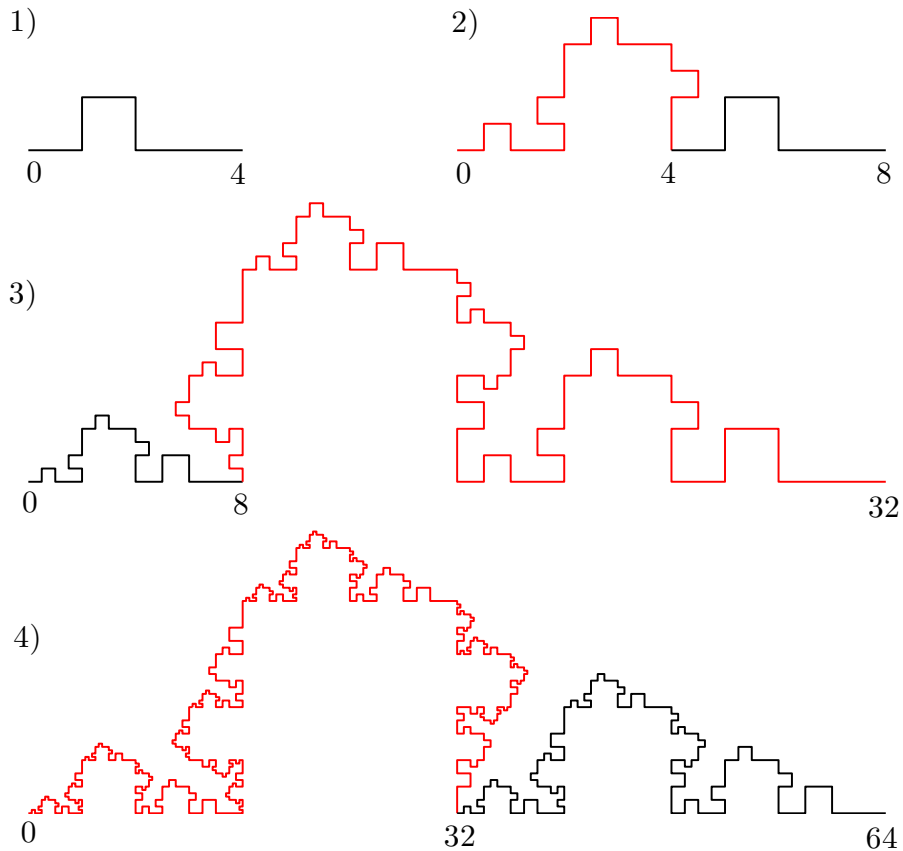}
\vspace{-1cm}                                             
\caption{\it Polygonals $p'_1$, $p'_2$, $p'_3$ and $p'_4$ of expansion process corresponding to Fig. \ref{fractal-ab}. Each new polygonal (in red) inherits the preceding one (in black) (steps {\rm 3)} and {\rm 4)} are out of scale).}
\label{ab-fractal}                                        
\end{figure}                                              

Hitherto, one ratio is chosen and inverted to make the expansion process in each step. But there is no reason for not using the inverse of some other ratio, or all of them, in the same process, selecting one for each iteration. 

For example, choose  the expansor $1/b\!=\!4$ for the odd steps $k$, and $1/a\!=\!2$ for the even ones. We take $p_1$ the generatrix of $F$ in Fig. \ref{fractal-ab}, and setting the origin at the left endpoint (only for clarity, not for necessity) we generate $p'_1$ four times longer than $p_1$ (Fig. \ref{ab-fractal}.1). This is equivalent, in this case, to stretching the interval $[0,1]$ to $[0,4]$, plus adding a ``square hump" keeping the shape of $p_1$. Next, we take $1/a\!=\!2$ as expansor, enlarging $p'_1$ by 2, but setting the origin at the right endpoint of $p'_1$ (again for simplicity), and adding square humps in a proportional way, obtaining $p'_2$ (Fig. \ref{ab-fractal}.2). Polygonal $p'_3$ is got from $p'_2$, $1/b\!=\!4$ the expansor, left endpoint the origin, adding the humps... and so on (Fig. \ref{ab-fractal}.3-4). There is inheritance: $p'_{k+1}$ contains $p'_k$, which guarantees the existence of a limit curve $\Gamma$. In this case, after $k$ expansions, we have the same number of segments and square humps that we would have by expanding only by $1/b$, and starting from the left, or by $1/a$, starting from the right. But, expanding $k$ times by $1/b$ would yield a ${p'}^b_k$ with a diameter $\Delta^b_k\!=\!(1/b)^k\!=\!4^k$. If we had proceeded using $1/a$, $\Delta^a_k$ would be $(1/a)^k\!=\!2^k$. Polygonals ${p'}^a_k$ and ${p'}^b_k$ would be identical in shape, but with different scales; with the same number of segments and humps, but of different diameters, $2^k$ and $4^k$: i.e., they would be similar. Instead, if we expand $k/2$ times by $1/a$ and $k/2$ times by $1/b$ {\bf no matter in which order}, we will obtain $p'_k$ with identical number of segments and humps as ${p'}^a_k$ and ${p'}^b_k$, but now with a diameter 
$$\Delta_k=(1/b)^{\frac{k}{2}}(1/a)^{\frac{k}{2}}=4^{\frac{k}{2}}2^{\frac{k}{2}}=\big((1/b)^{\frac{1}{2}}(1/a)^{\frac{1}{2}}\big)^{\,k}=\big(\!\sqrt{1\!/\!ab}\,\big)^{\,k}\ ,$$ 
a value strictly comprised between $1/a^k$ and $1/b^k$. This results exactly as if we had taken an expansor of $1\!/\!\sqrt{ab}$ in all iterations from the beginning, except that it would be impossible from a geometrical point of view, since there is no segment of length $\sqrt{ab}$ (or contractor $\sqrt{ab}\!<\!1$) in the generatrix $p_1$. This new limit curve, $\Gamma^{\sqrt{ab}}$, is half ``wrinkled" and half ``stretched", and thus has an intermediate dimension
$$\dim_{MF}(\Gamma^b)\ls\dim_{MF}\big(\Gamma^{\sqrt{ab}}\big)\ls\dim_{MF}(\Gamma^a)\ .$$
\subsubsection*{Remarks}
{\bf 1.} To start at the left (right) endpoint by $1/b$ ($1/a$) is arbitrary: it suffices to take any endpoint of any segment with length $b$ ($a$) to guarantee the polygonal $p'_k$ is fitted inside $p'_{k+1}$ after the enlargement, which ensures the inheritance.
\smallskip

{\bf 2.} Of course, one can expand making another choice, for example: one third of the time by $1/b$ and the remaining two third by $1/a$. In this case, for $k$ large, $\Delta_k\!\approx\!(1/b)^{\frac{1}{3}k}(1/a)^{\frac{2}{3}k}\!=\!\big(\!\sqrt[3]{1\!/\!a^2b}\big)^{\,k}$, with an expansor $c^{-1}\!=\!1\!/\!\sqrt[3]{a^2b}$, $c$ being strictly between $\sqrt{ab}$ and $a$. Thus, we have the following proposition.
\medskip

Given two contractors $a$ and $b$, we call $p^{a,b}$ the corresponding generatrix $p_1$ of $F$.

{\pro\label{continuization}\quad For all $c\!\in\!\mbb{R}$, $b\!<\!c\!<\!a$, there exists (up to a translation) a unique limit curve $\Gamma^c\!\in\!{\cal F}_{MF}$ constructed (not in a unique way) from $p^{a,b}$. Also
$$\dim_{MF}(\Gamma^b)\ls\dim_{MF}\big(\Gamma^c\big)\ls\dim_{MF}(\Gamma^a)\ .$$
}

\prf\quad Let $c$ be such that $b\!<\!c\!<\!a$. The function $g(x)\!=\!a\!\left(\!\dfrac{b}{a}\!\right)^x$ is strictly decreasing, with $g(0)\!=\!a$ and $g(1)\!=\!b$. Hence, there is a unique $\lambda$, $0\!<\!\lambda\!<\!1$ such that $g(\lambda)\!=\!c$, which satisfies
$$b^{\lambda}a^{1-\lambda}=c\ .$$
Let $\{r_k\}_{k\gs 1}$ be a sequence of natural numbers, $0\!\ls\!r_k\!\ls k$, such that $\ds\lim_{k\to +\infty}\dfrac{r_k}{k}\!=\!\lambda$ (e.g., $r_k\!=\![\lambda\!\cdot\!k]$). Then, we state an expansion process thus: in step $k$ we expand $r_k$ times by the factor $1/b$ , and $k\!-\!r_k$ times by the factor $1/a$. Clearly, there is no unique way of doing this. The diameter $\Delta_k$ in step $k$ will be 
$$\Delta_k=\left(\!\frac{1}{b}\!\right)^{\frac{\scrs r_k}{\scrs k}\,k}\!\left(\!\frac{1}{a}\!\right)^{\frac{\scrs (k-r_k)}{\scrs k}\,k}\sim\left(\!\frac{1}{b}\!\right)^{\lambda\,k}\!\left(\!\frac{1}{a}\!\right)^{(1-\lambda)\,k}=\left(\!\frac{1}{c}\!\right)^k\ ,$$
(where $\alpha_k\sim\beta_k$ means $\alpha_k/\beta_k\to 1$). 
Let $p^c_k$ be the corresponding polygonal, then $p^c_{k+1}$ inherits $p^c_k$ in a natural way, due to the construction, which guarantees the existence of a limit curve $\Gamma^c$. The proof of the inequalities of the dimensions is analogous and based on the same arguments used in Theorem \ref{dimMF-desig} proved in \cite{H-P1}.

\qed

In the general case of $N$ contractors $0\!<\!a_1\!\ls\!\cdots\!\ls\!a_N\!<\!1$, the proposition is valid, taking $a_i$ and $a_{i+1}$ instead of $b$ and $a$. Therefore, if $p^{a_1,\hdots,a_N}$ is the generatrix of $F$, we have
{\cor\quad For every $c_i\!\in\!\Bbb{R},\ a_i\!<\!c_i\!<\!a_{i+1}$, there exists a limit curve $\Gamma^{c_i}\!\in\!{\cal F}_{MF}$ constructed in terms of $p^{a_1,\hdots,a_N}$. Also
$$\dim_{MF}(\Gamma^{a_i})\ls\dim_{MF}\big(\Gamma^{c_i}\big)\ls\dim_{MF}(\Gamma^{a_{i+1}})\ .$$
}
So, every dimensional pair $\{\dim_{MF}(\Gamma^{a_i}),\dim_{MF}(\Gamma^{a_{i+1}})\}$ of the discrete MF spectrum is made a continuous dimensional interval $[\dim_{MF}(\Gamma^{a_i}),\dim_{MF}(\Gamma^{a_{i+1}})]$. Therefore
{\cor\label{continuization-2}\quad For any $c\!\in\!\Bbb{R},\ a_1\!<\!c\!<\!a_N$, there exists a limit curve $\Gamma^c\!\in\!{\cal F}_{MF}$ constructed from $p^{a_1,\hdots,a_N}$, and such that
$$\dim_{MF}(\Gamma^{a_1})\ls\dim_{MF}\big(\Gamma^c\big)\ls\dim_{MF}(\Gamma^{a_N})\ .$$}
We will call the continuous spectrum $[\dim_{MF}(\Gamma^{a_1}),\dim_{MF}(\Gamma^{a_N})]$, the \emph{MF spectrum} associated with $F$. From this corollary it is clear, and a remarkable point, that we do not need the intermediate contractors; only the minimal $a_1$ and the maximal $a_N$ are needed to make the discrete MF spectrum to be continuous.
Moreover, from Proposition \ref{continuization} and the Corollary \ref{continuization-2}, we have
{\cor \quad For every $a_i$, $i\!\neq\!1,N$, there exists $c,\ a_1\!<\!c\!<\!a_N$, and a curve $\Gamma^c$, stemming from expanding by $1/a_1$ and $1/a_N$ only, such that $\dim_{MF}(\Gamma^{a_i})\!=\!\dim_{MF}(\Gamma^c)$.
}
\medskip

\prf \quad Follows from Proposition \ref{continuization}, with $b\!=\!a_1$, $a\!=\!a_N$ and $c\!=\!a_i$.

\qed 
\section{Relations to Multifractal Spectra}
\subsection{Summary of basic notions}\label{basic notions}

The self similarity concept can be applied to measures. Let $S_1,\hdots,S_N$ be an IFS in $\mbb{R}^n$ with ratios 
$a_i\!\in\!(0,1)$, $1\!\ls\!i\!\ls\!N$. Recall that a Borel probability measure $\mu$ is called a self--similar measure (SSM) if
$$\mu(\cdot)=\sum^N_{i=1}\ p_i\,\mu(S^{-1}_i(\cdot))\ ,$$
where $p_i\!>\!0$, and $\sum_{i=1}^N \,p_i\!=\!1$. Hutchinson \cite{Hut} proved that such measures exist and are unique, and in this case $F\!=\!\supp(\mu)$, where $F$ is the invariant set for the IFS. Besides, it is known that, provided the IFS satisfies the OSC, all reasonable definitions of multifractal spectra of $\mu$ coincide \cite{Rie95,CM,Ols,Fal97}. They basically are: the `coarse' spectrum, $f_C$, related to box--counting dimension, the `fine' or singular spectrum, $f_H$ related to the Hausdorff one, and the Legendre transform $f_L$ \cite{Fal97}.

Briefly reviewing, should $F$ be covered by boxes $B_j$ of $\varep$--diameter, then for each $B_j$, we define
$$\alpha(B_j)=\frac{\log\mu(B_j)}{\log \varep}\ ,$$
and, for $\alpha\!>\!0$, let $N_\alpha$ be the number of boxes with $\alpha(B_j)\!\approx\!\alpha$, then 
$$f_C(\alpha)=\lim_{\varep\to 0}\frac{\quad \log N_\alpha}{-\log \varep}\ ,$$
if the limits exists.
Hausdorff spectrum $f_H$ is rather more related to a local concept: if $x\!\in\!F$, then
$$\alpha(x)=\lim_{\varep\to 0}\frac{\log\mu(B_x(\varep))}{\log \varep}\ ,$$
and
$$f_H(\alpha)=\dim_H\big\{x\in F: \alpha(x)=\alpha\big\}\ ,$$
where $B_x(\varep)$ is the ball of radius $\varep$ centered at $x$.

For $q\!\in\!\mbb{R}$ and $\varep\!>\!0$, consider 
\begin{equation}\label{tau-de-q}
\tau_\mu(q)=\lim_{\varep\to 0}\frac{\sum_j \mu(B_j)^q}{\log \varep}\ ,
\end{equation}
where the sum is over $\mu(B_j)\!>\!0$, for $B_j$ in an $\varep$--grid of $\mbb{R}^n$. Assuming the limit exists, $\tau_\mu(q)$ is the $L^q$--spectrum of $\mu$. 

The functions $f_C(\alpha)$ and $\tau(q)$ satisfy $\tau(q)\!=\!\inf_\alpha\{q\alpha\!-\!f_C(\alpha)\}$. Assuming differentiability of the functions (which is true for an SSM $\mu$), if for each $q$ the infimum is attained at $\alpha\!=\!\alpha(q)$, we have $q\!=\!f'_C(\alpha)$, $\tau(q)\!=\!q\alpha\!-\!f_C(\alpha(q))$ and $\alpha(q)\!=\!\tau'(q)$.

The Legendre transform $f_L$ of $\tau$ is, by definition
\begin{equation}\label{Legen-transf}
f_L(\alpha)=\inf_{-\infty<q<+\infty}\big\{q\alpha-\tau(q)\big\}\ .
\end{equation}
In this situation, $q$ and $\tau$ are related \cite{Lau} by 
\begin{equation}
\label{funcion-particion}
\sum_{i=1}^N\ p_i^q a_i^{-\tau(q)}=1\ ,
\end{equation}
called the \emph{partition function} due to its formal analogy with the partition function in statistical mechanics.

Thereafter, we will simply write $f\!:=\!f_C\!=\!f_H\!=\!f_L$.
Also related to the $L^q$--spectrum are the R\'enyi dimensions \cite{Ren}: let $\{B_j\}$ a partition of $\mbb{R}^n$ induced by an $\varep$--grid, and $\mu$ a probability measure supported on $F$, let $p_j\!=\!\mu(B_j)$. Then for $q\!\in\!\mbb{R}$, the R\'enyi spectrum of $\mu$ is, by definition
\renewcommand{\arraystretch}{2.5}
\begin{equation}
\label{dim-Renyi}
D_q(\mu)=\left\{
\begin{array}{ll}
\ds \lim_{\varep\to 0}\,\dfrac{1}{q\!-\!1}\dfrac{\log\left(\sum _j p_j^q\right)}{\log \varep}\ ,\quad &\quad {\rm para}\  q\neq 1 \ ,\\
\ds \lim_{\varep\to 0}\lim_{q\to 1}\,\dfrac{1}{q\!-\!1}\dfrac{\log\left(\sum_j p_j^q\right)}{\log \varep}\ ,\quad &\quad {\rm para}\  q=1 \ ,\\
\ds \lim_{\varep\to 0}\,\dfrac{\log (\sup_j p_j)}{\log \varep}\ ,\quad &\quad {\rm para}\  q=\!+\infty \ ,\\
\ds \lim_{\varep\to 0}\,\dfrac{\log (\inf_j p_j)}{\log \varep}\ ,\quad &\quad {\rm para}\  q=\!-\infty \ ,
\end{array}\right.
\end{equation}
\renewcommand{\arraystretch}{1.2}
By Eqs. (\ref{Legen-transf}) and (\ref{tau-de-q}) it is possible to relate the R\'enyi spectrum to multifractal $f(\alpha)$, and so to note that, for $q\!=\!0$, $D_0\!=\!f(\alpha(0))\!=\!\dim_{box}(F)$, for $q\!=\!1$, $D_1\!=\!f(\alpha(1))\!=\!\alpha(1)$, and for $q\!=\!+\infty$ and $q\!=\!-\infty$,
\begin{equation}
\label{D-infinity}
D_{+\infty}=\alpha(+\infty)=\alpha_{\rm min}\qquad {\rm and}\qquad D_{-\infty}=\alpha(-\infty)=\alpha_{\rm max}\ .
\end{equation}
In statistical mechanics, the partitions of a set are always considered of equal size. The $L^q$ spectrum $\tau(q)$ is, then, the free energy of the system described by $\mu$ as function of the inverse temperature $q$ (see for example \cite{Ott}). This corresponds to the case of having all the contractors $a_i$ being equal, which has been widely studied in the literature. In this case, the spectrum $(\alpha,f(\alpha))$ can be calculated in an explicit form \cite{Fal90, EM}. 

\subsection{Relation between the MF spectrum and \\ multifractal spectra}

Let us consider the partition of different size induced by the IFS yielding a curve $F\!\in\!{\cal F}_H$, and $\mu\!=\!(p_1,\hdots,p_N)$ an SSM on $F$. So, if $F_{i_1,\hdots,i_k}\!=\!S_{i_1}\!\circ\!\cdots\!\circ\! S_{i_k}(F)$, then $|F_{i_1,\hdots,i_k}|\!=\!a_{i_1}\!\cdots a_{i_k}\!=\!a_1^{r_1}\!\cdots a_N^{r_N}$ and $\mu(F_{i_1,\hdots,i_k})\!=\!p_{i_1}\!\cdots p_{i_k}\!=\!p_1^{r_1}\!\cdots p_N^{r_N}$, for $\sum_i^N r_i\!=\!k$. We will calculate the $f(\alpha)$ spectrum in terms of the ``frequencies" as the $\mu$ is distributed among the partitions.

For $x\!\in\!F$, let $F_k(x)$ the $k$--level set $F_{i_1,\hdots,i_k}$ that contains $x$, so we have 
\begin{equation}
\label{alpha-de-lambda}
\begin{array}{rcl}
\alpha=& \ds \lim_{k\to\infty}\frac{\log \mu(F_k(x))}{\log |F_k(x)|}=\lim_{k\to\infty}\frac{r_1\log p_1+\cdots+r_N\log p_N}{r_1\log a_1+\cdots+r_N\log a_N}&= \\
=&\!\ds  \frac{\lambda_1\log p_1+\cdots+\lambda_N\log p_N}{\lambda_1\log a_1+\cdots+\lambda_N\log a_N}\ =\ \frac{\log\big(p_1^{\lambda_1}\cdots p_N^{\lambda_N}\big)}{\log\big(a_1^{\lambda_1}\cdots a_N^{\lambda_N}\big)}\!&=\alpha(\lambda_1,\hdots,\lambda_N)\ ,
\end{array}
\end{equation}
provided the limits of frequencies $\lambda_i\!=\!\lim_{k\to \infty}r_i/k$ exist, $1\!\ls\!i\!\ls\!N$, where $r_i/k$ are the proportions of $p_i$ and $a_i$ in each $k$ step, so $\sum_i \lambda_i\!=\!1$. From this expression a known result for SSM can also be obtained
{\rem \label{alfa-min-max}\quad $\alpha_{\rm min}\!\leqslant\!\alpha\!\leqslant\!\alpha_{\rm max}$, where
$$\alpha_{\rm min}=\min_{1\leqslant i\leqslant N}\frac{\log p_i}{\log a_i}\qquad and \qquad \alpha_{\rm max}=\max_{1\leqslant i\leqslant N}\frac{\log p_i}{\log a_i}\ .$$}
\prf  \quad Indeed, it can be easily seen that the critical points of the function $\alpha(\lambda_1,\hdots,\lambda_N)$ subject to  $\sum_i \lambda_i\!=\!1$, are $(0,\hdots,i,\hdots,0)$, whose values of $\alpha$ are $\dfrac{\log p_i}{\log a_i}$, $1\!\ls\!i\!\ls\!N$.

\qed

The ubiquitous Stirling formula (as shown below) used in a standard way, yields  
$$f(\lambda_1,\hdots,\lambda_N)=\frac{\sum_i \lambda_i \log \lambda_i}{\sum_i \lambda_i \log a_i}\ .$$                                                                              

\bigskip

Let $F\!\in\!{\cal F}_H$ as above. Let $[d_{\min},d_{\max}]$ be the MF spectrum associated with $F$. In the remainder of this paper, we relate $[d_{\min},d_{\max}]$ to: {\bf(a)} the R\'enyi spectrum; {\bf(b)} the range of $f(\alpha)$ values, and {\bf (c)} the $[\alpha_{\min},\alpha_{\max}]$ range, choosing, in each case, an appropriate self--similar probability measure over $F$. In the first two cases, we will need to ``make" probabilities out of ratios. Since $\sum_i a_i\!=\!L\!>\!1$ we will need to ``contract the contractors", so that the new sum is unity. There are two ``natural" ways of doing it: $\sum_i a_i\!=\!L$ can be written as $\sum_i a_i/L\!=\!1$, and $a_i/L$ defined as the new $p_i$, which is what we do in Case  (a). The other way is obvious from the very symbol $d_H$, since $\sum_i a_i^{d_H}\!=\!1$, so $a_i^{d_H}$ would be the new $p_i$, that will be Case (b).
\medskip

{\bf Remark.} In case (a) we shrink each $a_i$ to $a_i/L$: it is, exactly, as if we zoom out generatrix $p_1$ (and $p_k$) until we see its length (from far away) to be 1 instead of $L\!>\!1$. In case (b) we shrink the $f(\alpha)$ spectrum, look at it from far away, until its height is 1, instead of $D_0\!>\!1$.
\subsubsection{Case (a)}

{\pro \quad Let $\ p_i\!=\!a_i/L$, then
$$d_{\min}=D_{+\infty}\qquad and \qquad d_{\max}=D_0\ .$$}

\prf \quad Indeed, by (\ref{D-infinity}), it suffices to show that $\alpha_{\min}\!=\!d_{\min}$
\begin{eqnarray}
\alpha_{\min}&=&\min_i\frac{\log p_i}{\log a_i}=\min_i\frac{\log (a_i/L)}{\log a_i}=\min_i \left\{1-\frac{\log L}{\log a_i}\right\}=\nonumber\\
               &=& \min_i \left\{1+\frac{\log L}{\log(1/a_i)}\right\}= 1+\frac{\log L}{\log(1/a_1)}\ .
\end{eqnarray}
So,  $\alpha_{\min}\!=\!\dim_{MF}(\Gamma^{a_1})\!=\!d_{\min}$.
\medskip

The other equality is true independently of the chosen probabilities $p_i$. Since $F$ is self--similar, it is known that $\dim_H(F)\!=\!\dim_{box}(F)$, then, it follows from Theorem \ref{dim-igual} that $d_{\max}\!=\!\dim_{MF}(\Gamma^{a_N})\!=\!D_0$.

\qed
\subsubsection{Case (b)}\label{case-b}

From Sec. \ref{basic notions}, we can write
$$F_\alpha=\left\{x\in F / \lambda(x)=(\lambda_1,\hdots,\lambda_N)=\lambda:\quad \frac{\sum_i\lambda_i\log p_i}{\sum_i\lambda_i\log a_i}=\alpha \right\}\ ,$$
for an SSM $\mu\!=\!(p_1,\hdots,p_N)$ chosen.
If $F_\lambda:=\{x\!\in\!F: \lambda(x)\!=\!\lambda\}$, then
\begin{equation}
\label{F-union-F}
\ds F_\alpha=\bigcup_{\lambda\,/ \alpha(\lambda)=\alpha}\! F_\lambda\ .
\end{equation}
\subsubsection*{(b.1) \mb{$N\!=\!2$}}

We have weights $p_1$ and $p_2$, and contractors $a_1$ and $a_2$: a binomial measure, $\lambda_1\!=\!\lambda$ and $\lambda_2\!=\!1\!-\!\lambda$. Therefore
$$\alpha=\alpha(\lambda)=\frac{\lambda \log p_1+(1\!-\!\lambda)\log p_2}{\lambda \log a_1+(1\!-\!\lambda)\log a_2}\ ,$$
so $F_\alpha\!=\!F_\lambda$, and $f(\alpha)\!=\!\dim_H(F_\alpha)\!=\!\dim_H(F_\lambda)$.
\medskip

For $k$, the weight of each segment $a_1^{r}a_2^{k-r}$ is $p_1^{r}\,p_2^{k-r}$, of which we have the binomial coefficient $C(k;r)$, i.e.
$$N_k(\alpha)= \binom{k}{r}\ .$$
Stirling's formula
$$k!\sim k^ke^{-k}\sqrt{2\pi k}\quad {\rm for}\quad k\to\infty\ ,$$
(where $a_k\!\sim\!b_k$ means $a_k/b_k\to 1$), allows us to write
$$\binom{k}{r}=\frac{k!}{r!\,(k\!-\!r)!}\sim  \frac{k^k}{r^r\,(k\!-\!r)^{\,k-r}}=\left(\left(\frac{r}{k}\right)^{\frac{r}{k}}\! \left(1\!-\!\frac{r}{k}\right)^{1\!-\!\frac{r}{k}} \right)^{-k}\ ,$$
$\lambda$ the frequency of $p$'s, and $1\!-\!\lambda$ that of $(1\!-\!p)$'s, so 
$$f(\alpha)=\lim_{k\to \infty}\frac{\log N_k(\alpha)}{-\log(a_1^ra_2^{k-r})}=\frac{\lambda\log \lambda+(1\!-\!\lambda)\log(1\!-\!\lambda)}{\lambda\log a_1+(1\!-\!\lambda)\log a_2}\ ,$$
i.e. $f(\alpha)=f(\lambda)$, a function of one parameter $0\!\ls\!\lambda\!\ls\!1$.
\subsubsection*{(b.2) \mb{$N\!>\!2$}}

Now, for the same value of $\alpha$ there are infinite vectors $\lambda\!=\!(\lambda_1,\hdots,\lambda_N)$ such that $\alpha(\lambda)\!=\!\alpha$. 
\medskip

For each $k$, we have $C(k;r_1,\hdots,r_N)$, the multinomial coefficient, $\sum_i r_i\!=\!k$, occurrences of weights $p_1^{r_1}\cdots p_N^{r_N}$ of $a_1^{r_1}\cdots a_N^{r_N}$, where 
$$\binom{k}{r_1,\hdots,r_N}=\frac{k!}{r_1!\cdots r_N!}\sim  \frac{k^k}{r_1^{r_1}\cdots r_N^{r_N}}=\left(\left(\frac{r_1}{k}\right)^{\frac{r_1}{k}}\cdots\left(\frac{r_N}{k}\right)^{\frac{r_N}{k}} \right)^{-k}\ ,$$
using Stirling's formula. Therefore
\begin{equation}
\label{f-alfa-frec}
f(\alpha(\lambda))\!=\!\frac{\lambda_1\log \lambda_1+\cdots+\lambda_N\log\lambda_N}{\lambda_1\log a_1+\cdots+\lambda_N\log a_N}\ .
\end{equation}
Since $f(\alpha(\lambda))\!=\!\dim_H(F_\lambda)$, then, $f(\alpha)\!=\!\sup\{f(\alpha(\lambda)):\alpha(\lambda)\!=\!\alpha\}$. 
\medskip

Thus, we will use Lagrange multipliers, and maximize the function
\begin{equation}
\label{maximizar}
h(\lambda)=f(\alpha(\lambda))\qquad \mathrm{subject\ to}\qquad \alpha(\lambda)=\alpha_0\ .
\end{equation}
Let $g$ be the auxiliary function, and $\Lambda\!\in\!\mbb{R}$
\begin{eqnarray}
g(\lambda,\Lambda)&=&h(\lambda)-\Lambda\,[\alpha(\lambda)-\alpha_0]= \label{fcion-aux} \\
          &=&\frac{\lambda_1\log \lambda_1+\!\cdots\!+\lambda_N\log\lambda_N}{\lambda_1\log a_1+\!\cdots\!+\lambda_N\log a_N}-\Lambda \left[ \frac{\lambda_1\log p_1+\!\cdots\!+\lambda_N\log p_N}{\lambda_1\log a_1+\!\cdots\!+\lambda_N\log a_N}- \alpha_0\right]\ .\nonumber
\end{eqnarray}
So, for $1\!\ls\!i\!\ls\!N$
\begin{eqnarray*}
0&=&\frac{\partial g}{\partial \lambda_i}=\\
&=&(\log\lambda_i\!+\!1\!)\!\sum_j\!\lambda_j\!\log a_j\!-\!\log a_i\!\sum_j\!\lambda_j\!\log\lambda_j\!-\!\Lambda\!\left[\!\log p_i\!\sum_j\!\lambda_j\!\log a_j\!-\!\log a_i\!\sum_j\!\lambda_j\!\log p_j\!\right]
\end{eqnarray*}
Then
$$0=\left(\sum_j\lambda_j\log a_j\right)(\log \lambda_i+1-\Lambda\log p_i)-\log a_i\left(\sum_j\lambda_j\log \lambda_j-\Lambda\sum_j\lambda_j\log p_j\right)\ ,$$
and dividing by $\sum_j\lambda_j\log a_j$ we obtain
$$0=\log \lambda_i+1-\Lambda\log p_i-\log a_i\left(\frac{\sum_j\lambda_j\log \lambda_j}{\sum_j\lambda_j\log a_j}-\Lambda\,\frac{\sum_j\lambda_j\log p_j}{\sum_j\lambda_j\log a_j}\right)\ .$$
Also
\begin{equation}
\label{restriccion}
\frac{\partial g}{\partial \Lambda}=0\qquad \mathrm{implies}\qquad 0=\alpha(\lambda)-\alpha_0\ .
\end{equation}
Now, for critical $\ol{\lambda}\!=\!(\ol{\lambda}_1,\!\hdots\!,\ol{\lambda}_N)$ we have, by (\ref{restriccion}) and (\ref{f-alfa-frec})
$$0=\log \ol{\lambda}_i+1-\Lambda\log p_i-\log a_i\Big(f(\alpha_0)-\Lambda \alpha_0 \Big)\qquad 1\!\ls\!i\!\ls\!N\ ,$$
and subtracting from this, the corresponding equation for $\lambda_1$, we have
$$0=\log\!\left(\!\frac{\ol{\lambda}_i}{\ol{\lambda}_1}\!\right)-\Lambda \log\!\left(\frac{p_i}{p_1}\!\right)- \log\!\left(\!\frac{a_i}{a_1}\!\right)\big(f(\alpha_0)-\Lambda \alpha_0 \big)\ ,$$
from which we obtain
\begin{equation}
\label{lambda-sub-i}
\ol{\lambda}_i=\ol{\lambda}_1\left(\!\frac{a_i}{a_1}\!\right)^{\!f(\!\alpha_0\!)-\Lambda\alpha_0}\!\left(\frac{p_i}{p_1}\!\right)^{\Lambda}\ ,
\end{equation}
and using $\sum_i\ol{\lambda}_i\!=\!1$ we have
$$1=\frac{\ol{\lambda}_1}{a_1^{f(\!\alpha_0\!)-\Lambda\alpha_0}\,p_1^\Lambda}\sum_i a_i^{f(\!\alpha_0\!)-\Lambda\alpha_0}\,p_i^\Lambda\ ,$$
from which 
$$\ol{\lambda}_1=\frac{a_1^{f(\!\alpha_0\!)-\Lambda\alpha_0}\,p_1^\Lambda}{\ds \sum_ia_i^{f(\!\alpha_0\!)-\Lambda\alpha_0}\,p_i^\Lambda}\ ,$$
and by (\ref{lambda-sub-i})
\begin{equation}
\label{lambdasub-i}
\ol{\lambda}_i=\frac{a_i^{f(\!\alpha_0\!)-\Lambda\alpha_0}\,p_i^\Lambda}{\ds \sum_j a_j^{f(\!\alpha_0\!)-\Lambda\alpha_0}\,p_j^\Lambda}\ ,
\end{equation}
for $i\!=\!1,\!\hdots\!,N$. Let $\Omega\!=\!\Omega(\Lambda)\!:=\!f(\alpha_0)\!-\!\Lambda\alpha_0$, so we can write
$$\log \ol{\lambda}_i = \log\big(a_i^\Omega\, p_i^\Lambda\big)-\log\!\left(\sum_ja_j^\Omega\, p_j^\Lambda \right)\ ,$$
calling $\bigstar\!:=\!\sum_ja_j^\Omega\,p_j^\Lambda$ (independent of $i$) for short, we multiply by $\ol{\lambda}_i$
$$\ol{\lambda}_i\log \ol{\lambda}_i=\ol{\lambda}_i \log\big(a_i^\Omega\, p_i^\Lambda\big)-\ol{\lambda}_i\log(\bigstar)\ ,$$
we add on $i\!=\!1,\!\hdots\!,N$, and then divide by $\sum_i \ol{\lambda}_i\log a_i$
$$\frac{\sum_i \ol{\lambda}_i\log \ol{\lambda}_i}{\sum_i \ol{\lambda}_i\log a_i}=\Omega\, \frac{\sum_i \ol{\lambda}_i\log a_i}{\sum_i \ol{\lambda}_i\log a_i}+\Lambda\,\frac{\sum_i \ol{\lambda}_i\log p_i}{\sum_i \ol{\lambda}_i\log a_i}-\frac{\log(\bigstar)}{\sum_i \ol{\lambda}_i\log a_i}\ ,$$
that is
$$f(\alpha_0)=\Omega+\Lambda\alpha_0 - \frac{\log(\bigstar)}{\sum_i \ol{\lambda}_i\log a_i}\ ,$$
therefore
$$\log(\bigstar)= 0\ ,$$
hence
\begin{equation}
\label{particion-2}
\sum_j a_j^{\Omega}\,p_j^\Lambda=\sum_j a_j^{f(\!\alpha_0\!)-\Lambda\alpha_0}\,p_j^\Lambda=1\ .
\end{equation}
Hence, from (\ref{lambdasub-i}) we obtain 
\begin{equation}
\label{lambdas-opt}
\ol{\lambda}_i=a_i^{\Omega}p_i^\Lambda=a_i^{f(\alpha_0)-\Lambda\alpha_0}p_i^\Lambda\ ,
\end{equation}
which we replace in (\ref{restriccion}) and obtain
\begin{equation}
\label{alfa-de-landon}
\alpha_0=\alpha(\ol{\lambda})=\alpha_0(\Lambda)=\frac{\ds \sum_i a_i^{\Omega}\,p_i^\Lambda\,\log p_i}{\ds \sum_i a_i^{\Omega}\,p_i^\Lambda\,\log a_i }
=\frac{\ds \sum_i a_i^{f(\!\alpha_0\!)-\Lambda\alpha_0}\,p_i^\Lambda\,\log p_i}{\ds \sum_i a_i^{f(\!\alpha_0\!)-\Lambda\alpha_0}\,p_i^\Lambda\,\log a_i}\ .
\end{equation}
We have then, for $\Lambda\!\in\!\mbb{R}$, a unique pair $(\alpha(\Lambda),f(\alpha(\Lambda)))$. If we are given the weights $p_1,\!\hdots\!,p_N$ then, once $\Lambda$ is fixed, $\sum_ja_j^{\Omega}\,p_j^\Lambda\!=\!1$ yield --numerically-- $\Omega(\Lambda)$, from which we obtain $\alpha(\Lambda)$ above, and  $f(\alpha(\Lambda))\!=\!\Omega(\Lambda)+\Lambda\alpha(\Lambda)$, or $f(\alpha)\!=\!\Omega(\Lambda)+\Lambda\alpha$. 
Since $\Lambda$ is the Lagrange multiplier, we have
$$\nabla h(\ol{\lambda})=\Lambda \nabla \alpha(\ol{\lambda})\ ,$$
and, by the very definition of function $h$ (\ref{maximizar}), we obtain
\begin{equation}
\label{multiplic=deriv}
\nabla h(\ol{\lambda})=f'(\alpha_0)\,\nabla\alpha(\ol{\lambda})\ ,
\end{equation}
and then
$$\Lambda=f'(\alpha_0)\ ,$$
provided $\nabla\alpha(\ol{\lambda})\!\neq\!0$. For each $i\!=\!1,\!\hdots\!,N$
\begin{eqnarray*} 
\frac{\partial\alpha}{\partial\lambda_i}(\ol{\lambda})&=&\frac{\partial}{\partial\lambda_i}\left(\frac{\sum_j\lambda_j\log p_j}{\sum_j\lambda_j\log a_j}\right)(\ol{\lambda})=\\
&=&\frac{\log p_i\sum_j\ol{\lambda}_j\log a_j-\log a_i\sum_j\ol{\lambda}_j\log p_j}{\left(\sum_j\ol{\lambda}_j\log a_j\right)^2}=\frac{\log p_i - \log a_i\ \alpha_0}{\sum_j\ol{\lambda}_j\log a_j}\ ,
\end{eqnarray*}
then
$$\nabla\alpha(\ol{\lambda})=0\quad \Longleftrightarrow\quad \alpha_0=\frac{\log p_i}{\log a_i}\quad \forall\ i\!=\!1,\!\hdots\!,N\ ,$$
that, by Remark \ref{alfa-min-max}, is only true if $\alpha(\lambda)$ is constant.
\medskip

{\bf Note.} The proof that $\ol{\lambda}$ is indeed a maximum for $h(\lambda)$ subject to $\alpha(\lambda)\!=\!\alpha_0$ is in the Appendix. Notice that $q\!=\!\dfrac{\mathrm{d}f}{\mathrm{d}\alpha}(\alpha(q))$ and $\tau(q)\!=\!q\alpha(q)\!-\!f(\alpha(q))$ identify $\Omega(\Lambda)\!=\!\Omega(f'(\alpha))\!=\!\Omega(q)\!=\!f(\alpha)\!-\!\Lambda\alpha$ as $-\tau(q)$, and 
$$\sum_ja_j^{\Omega}\,p_j^\Lambda=\sum_j\frac{p_j^q}{a_j^\tau}=1\ ,$$
is now the partition function.
\bigskip

Should the contractors be all equal, as in the case of the von Koch curve, we would have
$$\lambda_i=\frac{p_i^\Lambda}{\ds \sum_j p_j^\Lambda}\qquad i=1,\hdots,N\ ,$$
and the corresponding values of $\alpha(\Lambda)$ and $f(\alpha(\Lambda))$.
\medskip

If, instead, the probabilities are all equal, we have
\begin{equation}
\label{frec-prob-iguales}
\lambda_i=\frac{a_i^\Omega}{\ds \sum_j a_j^\Omega}\qquad i=1,\hdots,N\ ,
\end{equation}
and by (\ref{alfa-de-landon})
\begin{equation}
\label{alfa-prob-iguales}
\alpha=\alpha(\Omega)=\frac{\ds \sum_i\,\frac{a_i^\Omega}{\sum_j a_j^\Omega}\,\log\!\frac{1}{N}}{\ds \sum_i\,\frac{a_i^\Omega}{\sum_j a_j^\Omega}\,\log a_i}=\frac{\ds \log\!\frac{1}{N}}{\ds \sum_i\,\frac{a_i^\Omega}{\sum_j a_j^\Omega}\,\log a_i}\ ,
\end{equation}
and by (\ref{f-alfa-frec})
\begin{eqnarray}
\label{falfa-prob-iguales}
f(\alpha)=f(\alpha(\Omega))&=&\frac{\ds \sum_i\frac{a_i^\Omega}{\sum_ja_j^\Omega}\,\log\!\left(\!\frac{a_i^\Omega}{\sum_ja_j^\Omega}\!\right)}{\ds \sum_i\frac{a_i^\Omega}{\sum_ja_j^\Omega}\,\log a_i} =\nonumber\\
&=&\Omega-\frac{\log\!\left(\sum_ja_j^\Omega\right)}{\ds \sum_i\frac{a_i^\Omega}{\sum_ja_j^\Omega}\,\log a_i}\ .
\end{eqnarray}
We will choose this last case: all contractors have the same weight (lit. the same importance, the same value), no contractor is more significant than any other. This will allow us to give a thermodynamical interpretation to this case (b). So $p_i\!=\!1/N$, for $i\!=\!1,\hdots,N$, and we want to find, if possible, $\Omega_{\min}$ and $\Omega_{\max}$ such that 
$f(\alpha(\Omega_{\min}))\!=\!d_{\min}$ and $f(\alpha(\Omega_{\max}))\!=\!d_{\max}$.
\medskip

Writing $\Omega\!=\!D_0$ ($D_0\!=\!\dim_{box}(F)\!=\!\dim_H(F)$), by (\ref{frec-prob-iguales}) we have
$$\lambda_i=\frac{a_i^{D_0}}{\ds \sum_j a_j^{D_0}}= a_i^{D_0}\ ,\qquad i\!=\!1,\hdots,N\ ,$$
and by (\ref{falfa-prob-iguales})
$$f(\alpha(D_0))=D_0-\frac{\log\!\left(\sum_ja_j^{D_0}\right)}{\ds \sum_i\frac{a_i^{D_0}}{\sum_ja_j^{D_0}}\,\log a_i}=D_0\ ,
$$
for the value of $\alpha$
$$\alpha(D_0)=\frac{\ds \log\!\frac{1}{N}}{\ds \sum_ ia_i^{D_0}\log a_i}\ ,$$
according to (\ref{alfa-prob-iguales}). So, $f(\alpha(D_0))\!=\!d_{\max}\!=\!\dim_{MF}(\Gamma^{a_N})$, therefore $\Omega_{\max}\!=\!D_0$.
\medskip

Clearly, $\Omega_{\min}$ is the value of $\Omega$ satisfying $f(\alpha(\Omega))\!=\!d_{\min}$, that is
$$\Omega_{\min}-\frac{\log\!\left(\sum_ja_j^{\Omega_{\min}}\right)}{\ds \sum_i\frac{a_i^{\Omega_{\min}}}{\sum_ja_j^{\Omega_{\min}}}\,\log a_i}=d_{\min}\ ,$$
a value that cannot be known by analytical means. Yet, we can prove that it does exist. Indeed, let $0\!<\!a_1\!=\!\cdots\!=\!a_m\!<\!a_{m+1}\!\ls\!\cdots\!\ls\!a_N\!<\!1$, i.e. $m$ contractors equal to the smallest one, $1\!\ls\!m\!<\!N$. For short, let $f(\Omega)\!:=\!f(\alpha(\Omega))$. Then

{\lemma \quad $$\lim _{\Omega\to -\infty}f(\Omega)=\frac{\log(1/m)}{\log a_1}\ .$$}

\prf\quad If $\lambda_i(\Omega)\!:=\!\dfrac{a_i^\Omega}{\sum_j^N a_j^\Omega}=\dfrac{1}{\sum_j^N\left(\!a_j/a_i\!\right)^{\!\Omega}}\, ,$\quad then
$$\lim_{\Omega\to -\infty}\lambda_i(\Omega)=\left\{\begin{array}{cl} 1/m&\quad i=1,\hdots,m\ ,\\0 & \quad i=m\!+\!1,\hdots,N\ . \end{array}\right.$$
Hence, 
$$\lim _{\Omega\to -\infty}\lambda_i(\Omega)\log(\lambda_i(\Omega))=\left\{\begin{array}{cl} 1/m\,\log(1/m)&\quad i=1,\hdots,m\ ,\\0 & \quad i=m\!+\!1,\hdots,N\ . \end{array}\right.$$
Therefore, (\ref{falfa-prob-iguales}) yields
$$\lim _{\Omega\to -\infty}f(\Omega)=\lim _{\Omega\to -\infty} \frac{\sum_i\, \lambda_i(\Omega)\,\log(\lambda_i(\Omega))}{\sum_i\,\lambda_i(\Omega)\,\log a_i}=\dfrac{\log (1/m)}{\log a_1}\ .$$

\qed
\medskip

Since $f(\Omega)$ is continuous, and $f(D_0)\!=\!D_0\!=\!d_{\max}\!>\!d_{\min}$, then, the existence of $\Omega_{\min}$ ($\Omega_{\min}\!<\!D_0$), follows from $\dfrac{\log(1/m)}{\log a_1}<d_{\min}$. Let us recall that 
\begin{equation}
\label{dim-min}
d_{\min}\!=\!\dim_{MF}(\Gamma^{\ds a_1})= 1+ \dfrac{\log \ds\sum a_i}{\log\dfrac{1}{a_1}}=1+ \dfrac{\log L}{\log\dfrac{1}{a_1}}\ .
\end{equation}
Clearly
\begin{eqnarray*}
\dfrac{\log\dfrac{1}{m}}{\log a_1}= \dfrac{\log m}{\log\dfrac{1}{a_1}}&<& \frac{\log\!\left(\!\dfrac{a_1\!+\!\cdots\!+\!a_m\!+\!a_{m+1}\!+\!\cdots\!+\!a_N}{a_1}\!\right)}{\log\dfrac{1}{a_1}}=\\ &=&\frac{\log\dfrac{L}{a_1}}{\log\dfrac{1}{a_1}}\,=\,1+ \dfrac{\log L}{\log\dfrac{1}{a_1}}\,=\,d_{\min}\quad \surd
\end{eqnarray*}

Similarly, if $0\!<\!a_1\!\ls\!\cdots\!\ls\!a_{N-m'}\!<\!a_{N\!-m'+1}\!=\!\cdots\!=\!a_N\!<\!1$, i.e. $m'$ ratios are equal to $a_N$, then
{\cor\quad $$\lim _{\Omega\to +\infty}f(\Omega)=\frac{\log(1/m')}{\log a_N}\ .$$
}
\medskip

Since $p_i\!=\!1/N$, we have, by Remark \ref{alfa-min-max}
\begin{equation*}
\begin{array}{lc}
\ds \alpha_{\min}=\alpha(\Omega_{-\infty})=\min_{1\ls i\ls N}\frac{\log(1/N)}{\log a_i}=\frac{\log N}{\log(1/a_1)}\ ,& \mathrm{and}\\\\
\ds \alpha_{\max}=\alpha(\Omega_{+\infty})=\max_{1\ls i\ls N}\frac{\log(1/N)}{\log a_i}=\frac{\log N}{\log(1/a_N)}\ ,
\end{array}
\end{equation*}
therefore,
$$f(\alpha_{\min})=\frac{\log(1/m)}{\log a_1}\qquad \mathrm{and}\qquad f(\alpha_{\max})=\frac{\log(1/m')}{\log a_N}\ .$$
\medskip

We will compare $\Omega_{\min}$ with other values of $\Omega$ for which the corresponding $f(\alpha(\Omega))$ can be calculated.
\medskip

\begin{figure}[h]                                         
\includegraphics[width=17cm,height=12cm]{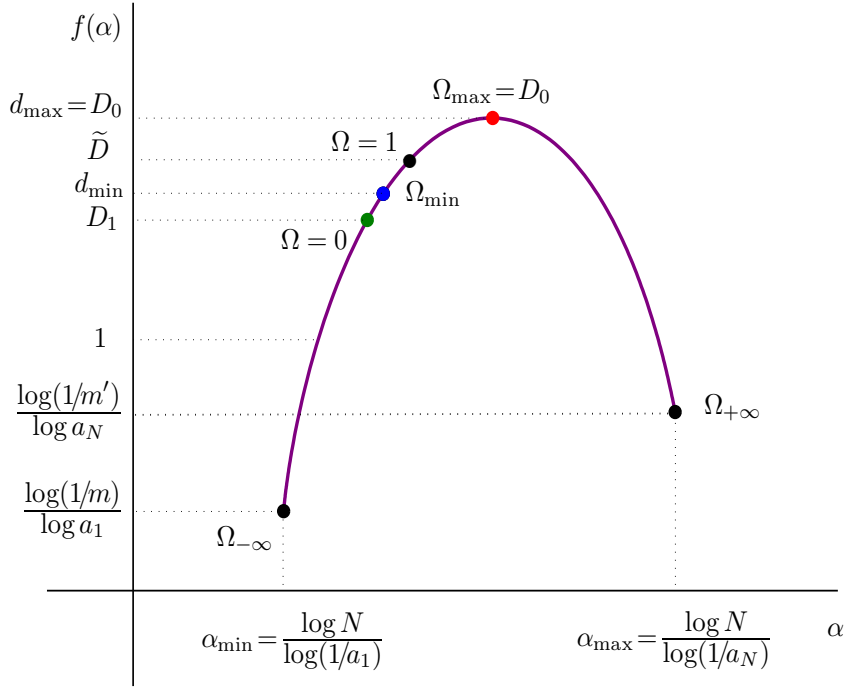}
\vspace{-3cm}                                             
\caption{\it Diagram of an $(\alpha(\Omega),f(\alpha(\Omega)))$ spectrum with its significant values ($d_{min}\!>\!D_1$ in this case). The MF spectrum covers the arc connecting $\dim_{MF}(\Gamma^{a_1})$ (blue point) to $\dim_{MF}(\Gamma^{a_N})$ (red point).}                         
\label{falfa-omegon}                                      
\end{figure}

With $\Omega\!=\!0$ we have, by (\ref{frec-prob-iguales})
$$\lambda_i=\frac{a_i^0}{\ds \sum_j a_j^0}= \frac{1}{N}\ ,\qquad i\!=\!1,\hdots,N\ ,$$
by (\ref{falfa-prob-iguales}) and (\ref{alfa-prob-iguales})
$$\alpha(0)=\frac{\ds \log\!\frac{1}{N}}{\ds \sum_ i \frac{1}{N}\log a_i}\ ,$$
and
\begin{equation}
\label{falfa-0}
f(\alpha(0))=0-\frac{\log\!\left(\sum_j a_j^0\right)}{\ds \sum_i\frac{a_i^0}{\sum_j a_j^0}\,\log a_i}=\frac{-\log N}{\ds \sum_ i \frac{1}{N}\log a_i}\ ,
\end{equation}
which means $f(\alpha(0))\!=\!\alpha(0)$, and $f(\alpha)\!=\!D_1$ (see Fig. \ref{falfa-omegon}).

Next, for $\Omega\!=\!1$ we have, by (\ref{frec-prob-iguales})
$$\lambda_i=\frac{a_i}{\ds \sum_j a_j}= \frac{a_i}{L}\ ,\qquad i\!=\!1,\hdots,N\ ,$$
by (\ref{falfa-prob-iguales}) and (\ref{alfa-prob-iguales})
$$\alpha(1)=\frac{\ds \log\!\frac{1}{N}}{\ds \sum_ i \frac{a_i}{L}\log a_i}\ ,$$
from which
\begin{equation}
\label{D-monio}
\wt{D}:=f(\alpha(1))=1-\frac{\log\!\left(\sum_j a_j\right)}{\ds \sum_i\frac{a_i}{\sum_j a_j}\,\log a_i}=1+\frac{\log L}{\ds \sum_ i \frac{a_i}{L}\log\dfrac{1}{a_i}}>1
\end{equation}
We claim: $d_{\min}\!\ls\!\widetilde{D}$, i.e. $\Omega_{\min}\!\ls\!1$.

Since (without loss of generality) $a_1\!\ls\!\cdots\!\ls\!a_N\!<\!1$, we have $1/a_i\!\ls\!1/a_1$ for $1\!\ls\!i\!\ls\!N$, hence
$$\sum_i \dfrac{a_i}{L}\log\dfrac{1}{a_i}\ls\sum_i \dfrac{a_i}{L}\log\dfrac{1}{a_1}=\log\dfrac{1}{a_1}\ds \sum_i \dfrac{a_i}{L}=\log\dfrac{1}{a_1}\ ,$$
so, the $d_{\min}\!\ls\!\widetilde{D}$ follows from  (\ref{dim-min}) and (\ref{D-monio}).
\medskip

Note that $D_1\!\ls\!d_{\min}$ is not always true: the situation depends on the choice of $a_i$. Let, for example, a generatrix of $F$ be
\begin{center}
\vspace{0.75cm}
\begin{picture}(180,20)
\put(0,0){\line(1,0){70}}\put(70,0){\line(2,3){20}}\put(110,0){\line(-2,3){20}}\put(110,0){\line(1,0){70}}
\put(32,5){$a$}\put(67,15){$b$}\put(107,15){$b$}\put(142,5){$a$}\put(0,-12){0}\put(178,-12){1}
\end{picture}
\vspace{0.75cm}
\end{center}

where, $N\!=\!4$, $a_1\!=\!a_4\!=\!a$ and $a_2\!=\!a_3\!=\!b$. $\sum_i a_i\!=\!L\!=2a+2b$ and $2a+b\!=\!1$. Let us write $b\!=\!a^p$, $p\!\gs\!1$  ($1/2\!\ls\!a\!\ls\!1/3$).
\medskip

For $p\!=\!2$, $2a+b\!=\!2a+a^2\!=\!1$, then $a\!=\!\sqrt{2}\!-\!1\!\approx\!0.41421$, $b\!=\!a^2\!=\!3\!-\!2\sqrt{2}\!\approx\!0.17157$. Hence
$$d_{\min}=1+\frac{\log L}{\log \dfrac{1}{b}}=1+\frac{\log(2a\!+\!2a^2)}{2\log \dfrac{1}{a}}\approx 1.08983\ .$$
whereas by (\ref{falfa-0})
$$D_1=f(\alpha(0))=\frac{-\log 4}{\dfrac{2}{4}\log a+\dfrac{2}{4}\log b}=\frac{-\log 4}{\dfrac{5}{4}\log a}\approx1.048585\ .$$
So, $D_1\!<\!d_{\min}$.
\medskip

But for $p\!=\!1.5$, $2a+b\!=\!2a+a^{1.5}\!=\!1$ ($a\!\approx\!0.382$, $b\!\approx\!0.236$), we have
$$d_{\min}\approx 1.147\qquad \mathrm{and} \qquad D_1\approx 1.152\ .$$
So, $d_{\min}\!<\!D_1$.

\subsection{Interpretation of parameter \mb{$\Omega$}}

\begin{figure}[h]                                             
\includegraphics[width=17cm,height=12cm]{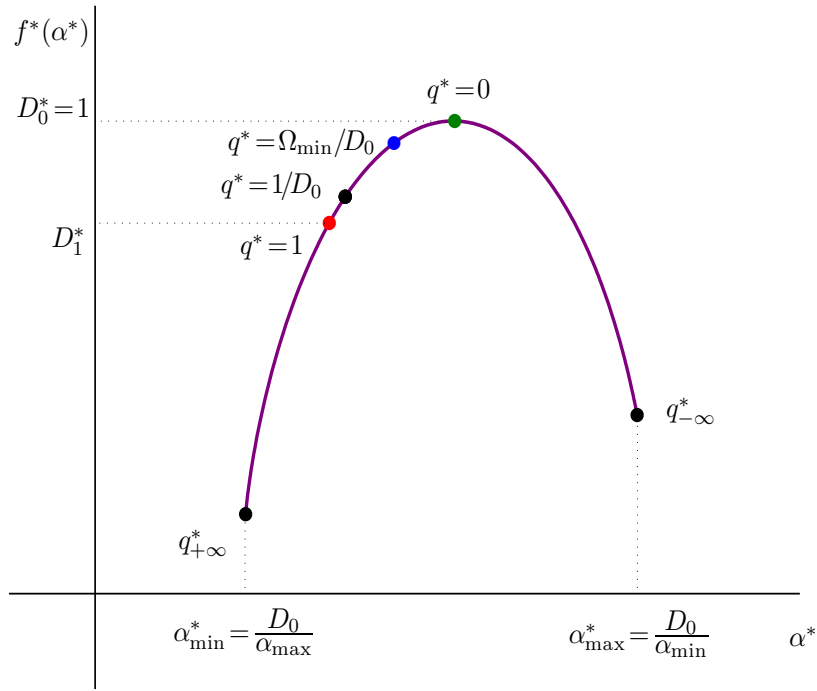}
\vspace{-3cm}                                                 
\caption{\it Diagram of inverted of shrunk spectrum $(\alpha^*,f^*(\alpha^*))$ of the one in Fig. \ref{falfa-omegon}. Here  $q^*$ can be interpreted as the slopes of tangents to the curve ($q^*\!=\!f^*(\alpha^*)$).}                           
\label{falfa-q-estrella}                                      
\end{figure}

The $(\alpha,f(\alpha))$ spectrum above, with $\sum _i (1/N)^q\, a_i^{-\tau}\!=\!1$, $-\tau(q)\!=\!\Omega$, $f'(\alpha)\!=\!q\!=\!\dfrac{\log\left(\sum_i a_i^{-\tau}\right)}{\log N}$ (and we have $\tau(q)$ given by $q(\tau)$), fulfills $\alpha_{\min}\!=\!\dfrac{\log 1/N}{\log a_1}$, $\alpha_{\max}\!=\!\dfrac{\log 1/N}{\log a_N}$, and $D_0\!=\!\max_\alpha f(\alpha)$. Now we shrink the spectrum, by shrinking both axes, horizontal and vertical, until $\max_\alpha f^{\rm sh}(\alpha)\!=\!1$, where $f^{\rm sh}$ is the shrunk spectrum. In case (a) we zoomed out the generatrix of $F$ until we ``saw" it to be of unit length instead of L>1; now we zoom out the spectrum until we ``see" its height to be unity, instead of $D_0$>1. As both axes rescale by $1/D_0$, we have now $\alpha^{\rm sh}_{\min}=\dfrac{\alpha_{\min}}{D_0}\!=\!\dfrac{\log 1/N}{D_0\log a_1}\!=\!\dfrac{\log 1/N}{\log a_1^{D_0}}$, and $\alpha^{\rm sh}_{\max}=\dfrac{\log 1/N}{\log a_N^{D_0}}$. We ``contracted the contractors" $a_i$ to a smaller $a_i^{D_0}$ value. Since $f$ and $f^{\rm sh}$ have exactly the same shape, the slopes of their tangents at any point have to coincide: $q\!=\!q^{\rm sh}$. Indeed we can write 
$$q=\frac{\log\!\left(\sum_ia_i^{-\tau}\right)}{\log N}=\dfrac{\log\!\left(\sum_i \left(a_i^{D_0}\right)^{-\tau/D_0}\right)}{\log N}=q^{\rm sh}\ ,$$
where $\tau$ has been replaced by $\tau/D_0$. The partition function above can be thus rewritten
$$1=\sum _i \dfrac{(1/N)^q}{a_i^\tau}=\sum_i \dfrac{(1/N)^{q^{\rm sh}}}{\left(a_i^{D_0}\right)^{\tau/D_0}}\ ,$$
with $q=q^{\rm sh}$ and $\tau$ replaced by $\tau/D_0$: this is the partition function for $f^{\rm sh}(\alpha)$. We have now $\sum_1^N 1/N=1=\sum_1^N a_i^{D_0}$, so we can invert contractors and probabilities \cite{RM}, obtaining a new spectrum: the inverted of the shrunk of the original $f(\alpha)$, which we call $f^*(\alpha)$. We have now $q^*=\dfrac{\mathrm{d}f^*(\alpha)}{\mathrm{d}\alpha}=-\dfrac{\tau}{D_0}$, using the permutation $q\!\leftrightarrow\!-\tau$ for inverse spectra, with $\tau/D_0$ in place of $\tau$. Now $-\tau=\Omega$, so the new derivative is $\Omega/D_0$: the inverse temperature for the entropy and internal energy given by $f^*(\alpha)$ and $\alpha$, according to the thermodynamical interpretation. The range $\Omega:1\to D_0$ is here $\Omega/D_0:1\to 1/D_0$: from the information or entropy dimension, up until slope $1/D_0$ (see Fig. \ref{falfa-q-estrella}).
\medskip

{\bf Note.} The same thermodynamical formalism applies to R\'enyi dimensions $D_q$: from their definition we have $(q-1)D_q=\tau(q)$, where $q$ is read as inverse temperature for the free energy $\tau(q)/q$. The abrupt change in the function $D_q$ at a certain value of $q$ is interpreted as a phase transition --at that $q$-- in many phenomena.

\subsubsection{Case (c)}

Let us now return to our $b\!<\!\sqrt{ab}\!<\!\sqrt[3]{a^2b}\!<\!a$ case above. We stressed that, though $b$ and $a$ were contractors, $\sqrt{ab}$ was not. It can be written as $a^\frac{1}{2}b^\frac{1}{2}$ and, expanding by $\ds c^{-1}\!=\!\frac{1}{\sqrt{ab}}$ we would have $p'_1$ as expanded by $\dfrac{1}{\sqrt{ab}}$, $p'_2$ as $p_2$ expanded by $\ds \left(\!\frac{1}{\sqrt{ab}}\!\right)^2\!=\!\frac{1}{ab}$, $p'_4$ would be $p_4$ expanded by $\ds \left(\!\frac{1}{ab}\!\right)^2$ (in any order: it could be $\dfrac{1}{a}\dfrac{1}{b^2}\dfrac{1}{a}\hdots$) etc. If we restrict ourselves to the {\bf even} $p'_k$, we have the very same limit curve: $\Gamma^{ab}$, where $ab$ would {\bf not} be a contractor in $p_1$, but in $p_2$: we would start from $p_2$ onwards (i.e. skipping the odd $p_k$ in the case $\ds c^{-1}\!=\!\frac{1}{\sqrt{ab}}$). With a suitable skipping we can obtain $\Gamma^{\sqrt[3]{a^2b}}$, which corresponds to $a^\frac{2}{3}b^\frac{1}{3}\hdots$ and the same is valid for any configuration where $c\!=\!a^{\lambda_a}b^{\lambda_b}$, $\lambda_a+\lambda_b\!=\!1$.

Now, in the general case of $a_1,\hdots,a_N$ contractors, let us choose a \emph{critical} vector $\ol{\lambda}\!=\!(\ol{\lambda}_1,\hdots,\ol{\lambda}_N)$ corresponding to a certain $\ds \alpha\!=\!\frac{\log 1/N}{\ol{\lambda}_1\log a_1\!+\!\cdots\!+\!\ol{\lambda}_N\log a_N}$, such that $f(\ol{\lambda})$ is precisely $f(\alpha)$, due to the maximization process in Subsec. \ref{case-b} (b.2). Since, by Eq. (\ref{alpha-de-lambda}) 
\begin{equation}
\label{critical-lambda}
\alpha=\frac{-\log N}{\ds \sum_i\ol{\lambda}_i \log a_i}=  \frac{\log N}{\log\!\bigg(\ds\frac{1}{a_1^{\ol{\lambda}_1}\cdots a_N^{\ol{\lambda}_N}}\!\bigg)}\ ,
\end{equation}
then, $\alpha$ growing implies $\log\!\big[1\big/\!\big(a_1^{\ol{\lambda}_1}\cdots a_N^{\ol{\lambda}_N}\big)\big]$ decreasing. Notice that $a_1^{\ol{\lambda}_1}\cdots a_N^{\ol{\lambda}_N}$, provided that $a_1^{\ol{\lambda}_i}$ are rational numbers, is a contractor $c$ that appears for the first time in a precise $p_{k_0}$ and, with a suitable skipping process, starting from $p'_{k_0}$ with expansor $c^{-1}\!=\!\big(a_1^{\ol{\lambda}_1}\cdots a_N^{\ol{\lambda}_N}\big)^{-1}$, generates the curve $\Gamma^{a_1^{\ol{\lambda}_1}\cdots a_N^{\ol{\lambda}_N}}$ (analogous to $\Gamma^{\frac{1}{2}\frac{1}{2}}$ above).

For instance, the $\ol{\lambda}$ critical for $\alpha_{\min}$ has the ``signature" $\big(\frac{1}{4},\frac{1}{4},\frac{1}{4},\frac{1}{4},0\big)$, which ensures only $b$'s and no $a$'s (i.e. $a_1$'s and no $a_2$'s). The signature for $ab$ (square of $\sqrt{ab}$) is $\big(\frac{1}{8},\frac{1}{8},\frac{1}{8},\frac{1}{8},\frac{1}{2}\big)$, which ensures half of $b$'s and half of $a$'s, exactly. The signature for $\alpha_{\max}$ is $(0,0,0,0,1)$, etc.

Let us fix a certain critical vector $\ol{\lambda}$, to which corresponds a certain (fixed) value of $\alpha$, we are in $p_k$ approaching $F$ (recall that any other than $\ol{\lambda}$ signature $(\lambda_1,\hdots,\lambda_N)$) yielding the same value of $\alpha$, would not produce $f(\lambda_1,\hdots,\lambda_N)\!=\!f(\alpha)$, but a smaller value, corresponding to $f(\alpha')$, $\alpha'\!\neq\!\alpha$. That is why we will work with critical signatures of $\ol{\lambda}$ for each $\alpha$).

Fixing $\alpha$ according to (\ref{critical-lambda}) implies fixing the length of segments $\big(a_1^{\ol{\lambda}_1}\cdots a_N^{\ol{\lambda}_N}\big)^k\!=\!a_1^{r_1}\cdots a_N^{r_N}$, $\sum_i r_i\!=\!k$, since we are in $p_k$ (for simplicity, we will refer to the values of $\alpha$ in polygonals $p_k$ approaching $F$). These segments of equal length approximate, as $k$ grows, an $\alpha$--Cantordust dense in $F$. The $\dim_H$ of this is precisely $f(\alpha)$. Going to the corresponding $\Gamma^{\ol{\lambda}}\!=\!\Gamma^{a_1^{\ol{\lambda}_1}\cdots a_N^{\ol{\lambda}_N}}$, $\ol{\lambda}\!=\!\ol{\lambda}(\alpha)$, we find in the corresponding nested $p'_k$ (with the same suitable skipping) segments of unit length, they are, exactly, those segments in $p_k$ of length $\big(a_1^{\ol{\lambda}_1}\cdots a_N^{\ol{\lambda}_N}\big)^k\!=\!a_1^{r_1}\cdots a_N^{r_N}$ when expanded by the $k$-th power of expansor $c^{-1}\!=\!\big(a_1^{\ol{\lambda}_1}\cdots a_N^{\ol{\lambda}_N}\big)^{-1}$. As $\alpha$ grows, the expansor $\big(a_1^{\ol{\lambda}_1}\cdots a_N^{\ol{\lambda}_N}\big)^{-1}$ decreases... and $\dim_{MF}(\Gamma^{\ol{\lambda}})$ increases, much as in the first example, when the expansor $1/b\!=\!1/a_1\!=\!4$ decreases from $k$--powers of 4 to $k$--powers of $1/a\!=\!2$... and $\dim_{MF}$ increases from $d_{\min}$ to $d_{\max}$. But then the number of unit segments in the $p'_k$ of each $\Gamma^{\ol{\lambda}}$, as $k$ grows, is a precise function of $\dim_{MF}(\Gamma^{\ol{\lambda}})$. Let us recall our simple example where $(a_1,a_2,a_3,a_4,a_5)\!=\!\big(\frac{1}{4},\frac{1}{4},\frac{1}{4},\frac{1}{4},\frac{1}{2}\big)$. Let us take $p_k$'s and $p'_k$'s with $k$ even, so $c\!=\!\sqrt{ab}$ appears as $c^2\!=\!ab$. If the underlying diameter is unity we have 16, 8, and 1 segments of length $1/16, 1/8$ and $1/4$ respectively. Now we expand by $(1/a)^2\!=\!2^2$, corresponding to $\alpha_{\min}$: we have 16, 8, and 1 segments of length $1/4,1/2$ and 1 respectively. Now let us expand by the other extreme value $\alpha_{\max}$, by $4^2$: we end up with 16, 8 and 1 segments of length 1, 2, and 4 respectively... And, by expanding by $\ds \frac{1}{a}\frac{1}{b}\!=\!c^{-1}\!=\!\left(\!\frac{1}{2}\frac{1}{4}\!\right)^{-1}\!=\!8$, an intermediate $\alpha$, we end up with 16, 8, and 1 segments of length $1/2, 1$ and 2 respectively: the number of unit segments in $p'_k$, as $k$ expands, is a give away of $\dim_{MF}(\ol{\lambda})$. (Remark: should the $\ol{\lambda}$ not be rational numbers, a limit process based on reasoning of Proposition \ref{continuization} would yield analogous results.)

We stress that we work with critical $\ol{\lambda}$ only, $\alpha(\ol{\lambda})$ yields a contractor $c\!=\!a_1^{\ol{\lambda}_1}\cdots a_N^{\ol{\lambda}_N}$ and an expansor $c^{-1}$, as evidenced by (\ref{critical-lambda}); $c\!=\!c(\ol{\lambda})$. To this contractor we have associated an $\alpha$--Cantordust, dense in $F$, its $\dim_H$ is, precisely, $f(\alpha)$. To the expansor $c^{-1}(\ol{\lambda})$ we have a family $p'_k$ converging to a $\Gamma^{\ol{\lambda}}\!\in\!{\cal F}_{MF}$. $\Gamma^{\ol{\lambda}}$ has unit segments given by expansor $c^{-1}$ applied to contractor $c$ in some $p_k$. There is a one--to--one connection thus described between $\alpha\!:\!\alpha_{\min}\to\alpha_{\max}$ with $\dim_{MF}\!:\!d_{\min}\to d_{\max}$. Two magnitudes grow: $\alpha$ and $\dim_{MF}$, as two others decrease: the expansor (from $4^k$ to $2^k$, in our simple example) and the number of unit segments in the corresponding $\Gamma$. The propagation of this number of unit segments in $p'_k$ as $k$ grows (propagation that can be quantified) is a fingerprint of the $\dim_{MF}$ of the corresponding $\Gamma$... But the quantitative law relating said propagation of unit segments to the corresponding $\dim_{MF}$ is, as yet, an open question.

\section{Conclusions}
Contractive processes producing a fractal bounded curve $F$ with Hausdorff dimension $\dim_H(F)\!>\!1$ can be associated with expansive processes producing locally smooth and unbounded curves $\Gamma$ with $\dim_{MF}(\Gamma)\!>\!1$. To each such $F$ (belonging to an ample family of curves) we associate a \MF\  MF dimensional spectrum, $\dim_{MF}\!\in\![d_{\min}, d_{\max}]$. Maximal $d_{\max}$ and minimal $d_{\min}$ of the MF spectrum are identified with Hausdorff and division dimensions of $F$. Such MF dimensional spectrum is compared with other multifractal spectra common in the literature, viz the spectrum of generalized R\'enyi dimensions, the thermodynamical formalism $(\alpha, f(\alpha))$, and a one--to--one correspondence between the $f(\alpha)$ dimensions and the $\dim_{MF}$ ones, via the critical frequency vectors $\ol{\lambda}$, which poses an open problem. The MF spectrum and the thermodynamical one are compared in terms of universal indices, together with their physical interpretation.

\setcounter{section}{0}
\renewcommand{\thesection}{\Alph{section}}
\section{Appendix: Maximality of \mb{$\ol{\lambda}$}}

To see that the critical frequencies $\ol{\lambda}_i$, $i\!=\!1,\hdots,N$, obtained in (\ref{lambdas-opt}), are indeed the maximum of $h(\lambda_1,\!\hdots\!,\lambda_N)$, we will use a classical tool from the theory of real valued functions of many variables: a determinant called the \emph{Bordered Hessian}, used for the second--derivative test in certain constrained optimization problems (\cite{MT91}): 

{\theo\label{extr-cond}  Let $h\!:\!U\!\subset\!\Bbb{R}^N\!\to\!\Bbb{R}$
and $\alpha\!:\!U\!\subset\!\Bbb{R}^N\!\to\!\Bbb{R}$ be of class ${\cal C}^2$. Let $\ol{\lambda}\!\in\!U$, $\alpha(\ol{\lambda})\!=\!\alpha_0$ and let $S$ be the level set of $\alpha$ of value $\alpha_0$. Suppose that $\nabla\alpha(\ol{\lambda})\!\neq\!0$ and that there exists a real number $\Lambda$ such that $\nabla h(\ol{\lambda})\!=\!\Lambda\nabla\alpha(\ol{\lambda})$. Let $g(\lambda)\!=\!h(\lambda)\!-\!\Lambda\alpha(\lambda)$ be the auxiliary function and the bordered Hessian

\begin{equation}
\label{hessiano}
|H_{N+1}|=\left|\begin{array}{ccccc}
0 & -\dfrac{\partial\alpha}{\partial\lambda_1} &-\dfrac{\partial\alpha}{\partial\lambda_2} & \cdots & -\dfrac{\partial\alpha}{\partial\lambda_N} \\ [0.5cm]
-\dfrac{\partial\alpha}{\partial\lambda_1} & \dfrac{\partial^2g}{\partial\lambda_1^2} & \dfrac{\partial^2g}{\partial\lambda_1\partial\lambda_2} & \cdots & \dfrac{\partial^2g}{\partial\lambda_1\partial\lambda_N} \\[0.5cm]
-\dfrac{\partial\alpha}{\partial\lambda_2} & \dfrac{\partial^2g}{\partial\lambda_1\partial\lambda_2} & \dfrac{\partial^2g}{\partial\lambda_2^2} & \cdots & \\[0.5cm]
\vdots &\vdots & \vdots &\ddots & \vdots \\[0.5cm]
-\dfrac{\partial\alpha}{\partial\lambda_N} & \dfrac{\partial^2g}{\partial\lambda_1\partial\lambda_N} & \dfrac{\partial^2g}{\partial\lambda_2\partial\lambda_N} & \cdots & \dfrac{\partial^2g}{\partial\lambda_N^2}
\end{array}\right|
\end{equation}
evaluated at $\ol{\lambda}$. Then, if the determinants of diagonal--submatrices of order $\geqslant\!3$ start with a subdeterminant $|H_3|\!>\!0$, and they alternate their signs ($|H_4|\!<\!0$, $|H_5|\!>\!0$,..., etc.), $\ol{\lambda}$ is a local maximum of $h$ constrained to $S$.
}

{\pro \quad The vector $\ol{\lambda}\!=\!(\ol{\lambda}_1,\!\hdots\!,\ol{\lambda}_N)$ obtained in (\ref{lambdas-opt}) is a maximum of $h(\lambda)$ (\ref{f-alfa-frec}) constrained to $\alpha(\lambda)\!=\!\alpha_0$, where $\alpha(\lambda)$ is the function form Eq. (\ref{alpha-de-lambda}).}
\medskip

\prf \quad Clearly, $h(\lambda)$ and $\alpha(\lambda)$ are functions of class ${\cal C}^2$ on $U\!\subset\!\Bbb{R}^N$, for 
$$U=\Big\{(\lambda_1,\hdots,\lambda_N):\ \lambda_i>0,\quad i\!=\!1,\!\hdots\!,N\Big\}\ ,$$ 
and $\ol{\lambda}\!=\!(\ol{\lambda}_1,\!\hdots\!,\ol{\lambda}_N)$ fulfills $\ol{\lambda}\!\in\!U$. We have also seen that $\nabla\alpha(\ol{\lambda})\!\neq\!0$, and for $\Lambda\!=\!f'(\alpha_0)$ (\ref{multiplic=deriv})
\begin{equation}
\label{grad-ptocrit}
\nabla h(\ol{\lambda})=\Lambda\nabla\alpha(\ol{\lambda})\ .
\end{equation}
\subsubsection*{Partial derivatives of \mb $\alpha(\lambda)$}
\begin{eqnarray}
\label{parc-alfa}
\frac{\partial}{\partial\lambda_i}\:\alpha(\lambda)&=&\frac{\partial}{\partial\lambda_i}\left(\frac{\sum_j\lambda_j\log p_j}{\sum_j\lambda_j\log a_j}\right)=\frac{\log p_i\sum_j\lambda_j\log a_j-\log a_i\sum_j\lambda_j\log p_j}{\left(\sum_j\lambda_j\log a_j\right)^{\!2}}= \nonumber\\ 
& = &\frac{\sum_j\lambda_j\log a_j}{\left(\!\sum_j\lambda_j\log a_j\!\right)^{\!2}}\left(\!\log p_i-\log a_i \frac{\sum_j\lambda_j\log p_j}{\sum_j\lambda_j\log a_j}\!\right)=\nonumber\\
&=& \frac{1}{\sum_j\lambda_j\log a_j}\Big(\!\log p_i-\log a_i\: \alpha(\lambda)\!\Big)\ .
\end{eqnarray}
Then
$$\frac{\partial\alpha}{\partial\lambda_i}(\ol{\lambda})= \frac{\log p_i-\log a_i\: \alpha_0}{\sum_j\ol{\lambda}_j\log a_j}\ .$$

\begin{eqnarray*}
\frac{\partial^2}{\partial \lambda_k\partial\lambda_i}\:\alpha(\lambda)&=&\frac{\partial}{\partial\lambda_k}\left[\!\frac{1}{\left(\!\sum_j\lambda_j\log a_j\!\right)}\Big(\!\log p_i-\log a_i\: \alpha(\lambda)\!\Big)\!\right]= \\
&=& \frac{-\log a_k}{\left(\!\sum_j\lambda_j\log a_j\!\right)^{\!2}}\Big(\!\log p_i-\log a_i\: \alpha(\lambda)\!\Big)+\frac{-\log a_i}{\left(\!\sum_j\lambda_j\log a_j\!\right)}\:\frac{\partial \alpha}{\partial \lambda_k}\ ,
\end{eqnarray*}
using (\ref{parc-alfa}), we obtain
\begin{equation}
\label{parc-ik-alfa}
\frac{\partial^2}{\partial \lambda_k\partial\lambda_i}\:\alpha(\lambda)= -\frac{\log a_k}{\left(\!\sum_j\lambda_j\log a_j\!\right)}\:\frac{\partial \alpha}{\partial \lambda_i}- \frac{\log a_i}{\left(\!\sum_j\lambda_j\log a_j\!\right)}\:\frac{\partial \alpha}{\partial \lambda_k}\ .
\end{equation}
In particular, for $k\!=\!i$
\begin{equation}
\label{parc-ii-alfa}
\frac{\partial^2}{\partial\lambda_i^2}\:\alpha(\lambda)= - \frac{2\:\log a_i}{\left(\!\sum_j\lambda_j\log a_j\!\right)}\:\frac{\partial \alpha}{\partial \lambda_i}\ .
\end{equation}
\subsubsection*{Partial derivatives of \mb $h(\lambda)$}
\begin{eqnarray}
\label{parc-h}
\frac{\partial}{\partial\lambda_i}\:h(\lambda)&=&\frac{\partial}{\partial\lambda_i}\left(\frac{\sum_j\lambda_j\log \lambda_j}{\sum_j\lambda_j\log a_j}\right)= \frac{(\log\lambda_i\!+\!1)\!\left(\!\sum_j\lambda_j\log a_j\!\right)-\log a_i\sum_j\lambda_j\log \lambda_j}{\left(\sum_j\lambda_j\log a_j\right)^{\!2}} =\nonumber \\
& = &\frac{\left(\!\sum_j\lambda_j\log a_j\!\right)}{\left(\!\sum_j\lambda_j\log a_j\!\right)^{\!2}}\left(\!(\log\lambda_i\!+\!1)-\log a_i \frac{\sum_j\lambda_j\log \lambda_j}{\sum_j\lambda_j\log a_j}\!\right)=\nonumber\\
&=& \frac{1}{\left(\!\sum_j\lambda_j\log a_j\!\right)}\Big(\!(\log\lambda_i\!+\!1)-\log a_i\: h(\lambda)\!\Big)\ .
\end{eqnarray}

For $k\!\neq\!i$
\begin{eqnarray*}
\frac{\partial^2}{\partial \lambda_k\partial\lambda_i}\:h(\lambda)&=&\frac{\partial}{\partial\lambda_k}\left[\!\frac{1}{\left(\!\sum_j\lambda_j\log a_j\!\right)}\Big(\!(\log\lambda_i\!+\!1)-\log a_i\: h(\lambda)\!\Big)\!\right]= \\
&=& \frac{-\log a_k}{\left(\!\sum_j\lambda_j\log a_j\!\right)^{\!2}}\Big(\!(\log\lambda_i\!+\!1)-\log a_i\: h(\lambda)\!\Big)+\frac{-\log a_i}{\left(\!\sum_j\lambda_j\log a_j\!\right)}\:\frac{\partial h}{\partial \lambda_k}\ ,
\end{eqnarray*}
using (\ref{parc-h}), we obtain
\begin{equation}
\label{parc-ik-h}
\frac{\partial^2}{\partial \lambda_k\partial\lambda_i}\:h(\lambda)= -\frac{\log a_k}{\left(\!\sum_j\lambda_j\log a_j\!\right)}\:\frac{\partial h}{\partial \lambda_i}- \frac{\log a_i}{\left(\!\sum_j\lambda_j\log a_j\!\right)}\:\frac{\partial h}{\partial \lambda_k}\ .
\end{equation}
For $k\!=\!i$
\begin{eqnarray*}
\frac{\partial^2}{\partial \lambda_i^2}\:h(\lambda)&=&\frac{\partial}{\partial\lambda_i}\left[\!\frac{1}{\left(\!\sum_j\lambda_j\log a_j\!\right)}\Big(\!(\log\lambda_i\!+\!1)-\log a_i\: h(\lambda)\!\Big)\!\right]= \\
&=& \frac{-\log a_i}{\left(\!\sum_j\lambda_j\log a_j\!\right)^{\!2}}\Big(\!(\log\lambda_i\!+\!1)-\log a_i\: h(\lambda)\!\Big)+\frac{1}{\left(\!\sum_j\lambda_j\log a_j\!\right)}\left(\!\frac{1}{\lambda_i}-\log a_i\:\frac{\partial h}{\partial \lambda_i}\!\right)\ ,
\end{eqnarray*}
again, by (\ref{parc-h}) 
\begin{equation}
\label{parc-ii-h}
\frac{\partial^2}{\partial\lambda_i^2}\:h(\lambda)= \frac{1/\lambda_i}{\left(\!\sum_j\lambda_j\log a_j\!\right)}-\frac{2\,\log a_i}{\left(\!\sum_j\lambda_j\log a_j\!\right)}\:\frac{\partial h}{\partial \lambda_i}\ .
\end{equation}
\subsubsection*{Second--order partial derivatives of \mb $g(\lambda)$}

For $k\!\neq\!i$, $g(\lambda)\!=\!h(\lambda)\!-\Lambda\alpha(\lambda)$, then, by (\ref{parc-ik-alfa}) and (\ref{parc-ik-h})
\begin{eqnarray*}
\label{parc-ik-g}
\frac{\partial^2g}{\partial \lambda_k\partial\lambda_i}&=&\frac{\partial^2h}{\partial \lambda_k\partial\lambda_i}-\Lambda\frac{\partial^2\alpha}{\partial \lambda_k\partial\lambda_i}=\\
&=&  -\frac{\log a_k}{\left(\!\sum_j\lambda_j\log a_j\!\right)}\left[\frac{\partial h}{\partial \lambda_i}-\Lambda\frac{\partial \alpha}{\partial \lambda_i} \right] -\frac{\log a_i}{\left(\!\sum_j\lambda_j\log a_j\!\right)}\left[\frac{\partial h}{\partial \lambda_k}-\Lambda\frac{\partial \alpha}{\partial \lambda_k} \right]\ , 
\end{eqnarray*}
therefore, from (\ref{grad-ptocrit})
\begin{equation}
\label{g-ik-cero}
\frac{\partial^2g}{\partial\lambda_k\partial\lambda_i}(\ol{\lambda})=0\ .
\end{equation}
For $k\!=\!i$, from (\ref{parc-ii-alfa}) and (\ref{parc-ii-h})
\begin{eqnarray*}
\label{parc-ii-g}
\frac{\partial^2}{\partial\lambda_i^2}\:g(\lambda)&=&\frac{\partial^2}{\partial\lambda_i^2}\:h(\lambda)-\Lambda\frac{\partial^2}{\partial\lambda_i^2}\:\alpha(\lambda)=\\
&=& \frac{1/\lambda_i}{\left(\!\sum_j\lambda_j\log a_j\!\right)}-\frac{2\ \log a_i}{\left(\!\sum_j\lambda_j\log a_j\!\right)}\left[\frac{\partial h}{\partial \lambda_i}-\Lambda\frac{\partial \alpha}{\partial \lambda_i} \right] \ , 
\end{eqnarray*}
then, from (\ref{grad-ptocrit})
\begin{equation}
\label{g-ii}
\frac{\partial^2g}{\partial\lambda_i^2}(\ol{\lambda})=\frac{1/\ol{\lambda}_i}{\sum_j\ol{\lambda}_j\log a_j}<0\ .
\end{equation}
\bigskip

Next, writing $A_i\!=\!\dfrac{\partial\alpha}{\partial\lambda_i}(\ol{\lambda})$ and $B_i\!=\!\dfrac{\partial^2g}{\partial\lambda_i^2}(\ol{\lambda})$, the bordered Hessian (\ref{hessiano}) evaluated at $\ol{\lambda}$, is
\begin{equation}
\label{hess-g}
|H_{N+1}|=\left|\begin{array}{ccccc}
0 & -A_1 &-A_2 & \cdots & -A_N \\ [0.2cm]
-A_1 & B_1 & 0 & \cdots & 0 \\[0.2cm]
-A_2 & 0 & B_2 & \cdots & 0 \\ [0.2cm]
\vdots &\vdots & \vdots &\ddots & \vdots \\ [0.2cm]
-A_N & 0 & 0 & \cdots & B_N
\end{array}\right|
\end{equation}
Notice that, since $\nabla\alpha(\ol{\lambda})\!\neq\!0$, it can be assumed, without loss of generality, that $A_1\!\neq\!0$. Then, for $n\!=\!2$ 
\begin{eqnarray*}
|H_3|&=&\left|\begin{array}{ccc} 0 & -A_1 & -A_2 \\ -A_1 & B_1 & 0\\ -A_2 & 0 & B_2\end{array}\right|=B_2 \left|\begin{array}{cr} 0 & -A_1\\ -A_1 & B_1\end{array}\right|+(-A_2)\left|\begin{array}{cc} -A_1 & B_1 \\ -A_2 & 0\end{array}\right|= \\ [0.2cm]
&=& -B_2 A_1^2-B_1A_2^2=B_2|H_2|-B_1A_2^2\ ,
\end{eqnarray*}
hence, from (\ref{g-ii})
$$|H_3|>0\ .$$
It can be easily seen that, for $3\!\ls\!n\!\ls\!N$ we have
\begin{eqnarray*}
|H_{n+1}|&=&B_n |H_n|+(-1)^{n-1}A_n^2 \left|\begin{array}{ccc} B_1 & \cdots & 0 \\ \vdots & \ddots&\vdots\\ 0 & \cdots & B_{n-1}\end{array}\right|= \\
&=& B_n |H_n|+(-1)^{n-1}A_n^2\prod_{i=1}^{n-1}B_i\ ,
\end{eqnarray*}
and, by a recurring process  
$$\mathrm{sgn}(|H_n|)=(-1)^{n+1} \ .$$
Hence, from Theorem \ref{extr-cond}, the vector $\ol{\lambda}$ maximizes $h(\lambda)$ constrained to $\alpha(\lambda)\!=\!\alpha_0$.

\qed
\section*{Acknowledgments}
This work was partially supported by UBACyT, Project I-420 (2008-2010), Ministerio de Educaci\'on, Argentina.



\begin{thebibliography}{xx}

\bibitem[1]{Man82} Mandelbrot, B., {\it The Fractal Geometry of Nature}, Freeman and Co., San Francisco, (1982).

\bibitem[2]{MF} Mend\`es France, M., ``The Planck Constant of a Curve", {\it Fractal Geometry and Analysis}, Kluwer Academic Publishers,  pp. 325--366, (1991).

\bibitem[3]{Stz} Strichartz, R., ``Fractals in the Large", {\it Canadian Journal of Mathematics}, Vol. 50, {\bf 3}, pp. 638--657, (1998).

\bibitem[4]{CT} Tricot, C., {\it Curves and Fractal Dimension}, Springer--Verlag, New York, (1995).

\bibitem[5]{Fal90} Falconer, K., {\it Fractal Geometry. Mathematical Foundations and Applications}, John Wiley \& Sons, (1990).

\bibitem[6]{Fal97} Falconer, K., {\it Techniques in Fractal Geometry}, John Wiley \& Sons, (1997).

\bibitem[7]{H-P1} Hansen, R., Piacquadio, M., ``The Dimension of Hausdorff and Mend\`es France. A Comparative Study",  {\it Revista de la Uni\'on Matem\'atica Argentina}, Vol. 42, {\bf 2} , pp.17--33, (2001). 

\bibitem[8]{Hut} Hutchinson, J.E., ``Fractals and Self-similarity", \emph{Indiana University Mathematics Journal}, {\bf 30}, pp. 713--747, (1981).

\bibitem[9]{Rie95} Riedi, R., ``An Improved Multifractal Formalism and Self-Similar Measures", {\it Journal of Math. Analysis and Applications}, {\bf 189}, pp. 462--490, (1995).

\bibitem[10]{CM} Cawley, R., Mauldin, R.D., ``Multifractal Decompositions of Moran Fractals", {\it Advances in Mathematics}, {\bf 92}, pp. 196--236, (1992).

\bibitem[11]{Ols} Olsen, L., ``A Multifractal Formalism", {\it Advances in Math.}, {\bf 116}, pp. 82--196, (1995).

\bibitem[12]{Lau} Lau, K-S., ``Self--similarity $L^q$--spectrum and Multifractal Formalism", {\it Progress in Probability}, {\bf 37}, pp. 55--90, (1995).

\bibitem[13]{Ott} Ott, E., {\it Chaos in Dynamical Systems}, Cambridge University Press, (1993).

\bibitem[14]{EM} Evertsz, C.J., Mandelbrot, B.B., ``Multifractal Measures",  Appendix B in: {\it Chaos and Fractals} by  H.O. Peitgen, H. Jürgens and D. Saupe, Springer New York, pp. 849--881, (1992).

\bibitem[15]{Ren} R\'enyi, A., ``On the Dimension and Entropy of Probability Distributions", {\it Acta Math. Acad. Sci. Hung.}, {\bf 10}, pp. 193--215, (1959).

\bibitem[16]{RM} Riedi, R., Mandelbrot, B., ``The Inversion Formula for Continuous Multifractals", {\it Advances in Applied Mathematics}, {\bf 19}, pp. 332--354, (1997).

\bibitem[17]{MT91} Marsden, J., Tromba, A., {\it Cálculo Vectorial}, Addison--Wesley Iberoamericana (tercera edición), (1991).



\end{thebibliography}
\end{document}